\documentclass{aa}  

\usepackage{graphicx}
\usepackage{setspace}
\usepackage{natbib}
\usepackage{rotating}
\usepackage{threeparttable}
\usepackage{tabto}
\usepackage{float}
\usepackage{subcaption}
\usepackage{caption}
\usepackage{amsmath}
\usepackage{nccmath}
\bibpunct{(}{)}{;}{a}{}{,} 

\usepackage{txfonts}
\usepackage[colorlinks,pagebackref,pdfusetitle,urlcolor=blue,citecolor=blue,linkcolor=blue,bookmarksnumbered,plainpages=false]{hyperref}
\begin{document} 

   \title{Properties of luminous red supergiant stars \\in the Magellanic Clouds}

   \author{S. de Wit
          \inst{1,2} \and A.Z. Bonanos \inst{1} \and F. Tramper \inst{1,3} \and M. Yang \inst{1,4} \and G. Maravelias \inst{1,5} \and K. Boutsia \inst{6} \and N. Britavskiy \inst{7} \and E. Zapartas \inst{1,8}
          }

   \institute{
    IAASARS, National Observatory of Athens, 15326 Penteli, Greece
    \and
    National and Kapodistrian University of Athens, Department of Physics, Panepistimiopolis, Zografos, GR15784, Greece 
    \and
    Institute of Astronomy, KU Leuven, Celestijnlaan 200D, 3001 Leuven, Belgium
    \and
    Key Laboratory of Space Astronomy and Technology, National Astronomical Observatories, Chinese Academy of Sciences, Beijing 100101, People’s Republic of China
    \and
    Institute of Astrophysics FORTH, GR-71110, Heraklion, Greece
    \and
    Carnegie Observatories, Las Campanas Observatory, Colina El Pino, Casilla 601, La Serena, Chile
    \and
    University of Li\`ege, All\'ee du 6 Ao\^ut 19c (B5C), B-4000 Sart Tilman, Li\`ege, Belgium
    \and
    Department of Astronomy, University of Geneva, Chemin Pegasi 51, CH-1290 Versoix, Switzerland}

   \date{}
   
 
  \abstract
   {There is evidence that some red supergiants (RSGs) experience short lived phases of extreme mass loss, producing copious amounts of dust. These episodic outburst phases help to strip the hydrogen envelope of evolved massive stars, drastically affecting their evolution. However, to date, the observational data of episodic mass loss is limited.}
   {This paper aims to derive surface properties of a spectroscopic sample of fourteen dusty sources in the Magellanic Clouds using the Baade telescope. These properties may be used for future spectral energy distribution fitting studies to measure the mass loss rates from present circumstellar dust expelled from the star through outbursts.}
   {We apply \textsc{marcs} models to obtain the effective temperature ($T_{\rm eff}$) and extinction ($A_V$) from the optical TiO bands. We use a $\chi^2$ routine to determine the best fit model to the obtained spectra. We compute the $T_{\rm eff}$ using empirical photometric relations and compare this to our modelled $T_{\rm eff}$.}
   {We have identified a new yellow supergiant and spectroscopically confirmed
  eight new RSGs and one bright giant in the Magellanic Clouds. Additionally, we observed a supergiant B[e] star and found that the spectral type has changed compared to previous classifications, confirming that the spectral type is variable over decades. For the RSGs, we obtained the surface and global properties, as well as the extinction $A_V$.}
   {Our method has picked up eight new, luminous RSGs. Despite selecting dusty RSGs, we find values for $A_V$ that are not as high as expected given the circumstellar extinction of these evolved stars. The most remarkable object from the sample, LMC3, is an extremely massive and luminous evolved massive star and may be grouped amongst the largest and most luminous RSGs known in the Large Magellanic Cloud (log(L$_*$/L$_{\odot})\sim$5.5 and $R = 1400 \,\ \textrm R_{\odot}$).}

   \keywords{stars: massive - stars: supergiants - stars: fundamental parameters - stars: atmospheres - stars: late-type - Magellanic Clouds 
}

\maketitle

\section{Introduction}

The red supergiant (RSG) phase is the last evolutionary phase for the majority of massive stars \citep{Levesque2017}. RSGs are subject to intense mass loss through stellar winds and outbursts \citep[i.e. short lived phases of extreme mass loss, see][for a review]{Smith2014}. These outbursts remain poorly constrained, despite increasing evidence supporting their existence \citep{Prieto2008,Mauerhan2013,Bruch2021, Humphreys2022}. The ability to potentially strip massive stars of their hydrogen envelope may drastically affect the course of evolution of massive stars in the latest stages before supernova collapse. At present, stellar evolutionary models for single stars \citep[e.g.][]{Eggenberger2008,Eggenberger2021,Brott2011a,Ekstrom2012,Kohler2015} or for binary stars \citep[e.g. BPASS;][]{BPASS1} adopt a continuous mass-loss rate, strongly influencing the type of supernova event. One approach to understand episodic mass loss in evolved massive stars, is to study the stellar properties of these stars empirically and test whether the derived properties can be recovered by existing evolutionary predictions. Several recent works have studied the mass loss properties \citep[e.g.][]{Beasor2020,Beasor2021} and stellar properties \citep[e.g.][]{Massey2021,Gonzalez2021} of stars in the RSG phase. Interestingly, \cite{Beasor2020,Beasor2021} have found the mass-loss rates in the RSG phase to be lower than that of classical recipes \citep[i.e.][]{Jager1988,Loon2005}, predicting that most single stars fail to remove their hydrogen rich envelope and therefore should explode as a type IIP supernova. This is inconsistent with the lack of observed high mass type IIP progenitor stars, known as the "red supergiant problem" \citep{Smartt2009,Smartt2015,Daveis2020}. The solution to this contradiction may either reside in poorly understood short-lived phases \citep[$\sim\!\!10^4$ yrs;][]{Beasor2020} of extreme mass loss, in addition to outbursts in the last years before the supernova explosion, resulting in a type IIn supernova \citep{Smith2009,Smith2014} or in stripping of the envelope due to binary interactions \citep{Podsiadlowski1992,Eldridge2013,Zapartas2017}. \\
\indent
The mass-loss rates of RSGs can be measured using SED fitting techniques \citep[e.g. DUSTY;][]{Dusty1}. RSGs can reveal an excess emission due to circumstellar dust as a result of slow and thick stellar winds in the RSG phase. When an object is surrounded by circumstellar dust, one can derive the contribution from the central component and the dust component by fitting the shape of the full SED \citep[][Yang et al. 2022, in prep.]{Goldman2017}. Reliable mid-infrared (IR) photometry to model the dust emission bumps \citep[i.e.\ 10$\mu$m $\&$ 18$\mu$m silicate emission and 11.3$\mu$m PAH emission;][]{Jones2017} is vital, but such data are scarce due to instrument limitations, especially for more distant galaxies. Constraining the properties of the central source through spectroscopy eliminates potential degeneracies and uncertainties in the SED fitting procedure, greatly reducing the number of viable models for fitting the SED. To assess the mass loss, we therefore first need to derive accurate stellar parameters, such as the effective temperature of the central source. We approach this by targeting RSGs in the Large Magellanic Cloud (LMC) and Small Magellanic Cloud (SMC), using selection criteria from the near-IR and mid-IR \citep[e.g.][]{Bonanos2009,Bonanos2010,Britavskiy2014}, to look for potential dust-obscured RSGs. In recent decades, various methods have been applied to derive the properties of RSGs. Using the \textsc{marcs} model atmospheres \citep{Gustafsson2008}, \cite{Davies2013} and \cite{Gonzalez2021} have fitted the shape of the SED, while \cite{Davies2015} and \cite{Tabernero2018} have used these models to fit spectral lines in the $J$ and $I-$band, respectively. In our work however, as we had access to optical spectra, we used a similar approach to \cite{Levesque2005,Levesque2006}, where we measure the effective temperature from the depths of the optical TiO bands.\\
\indent
In Section \ref{Sec2} we discuss the selection of targets and describe the spectroscopic observations, data reduction and spectral classification of the spectra. We present the results of spectral modelling in Section \ref{Sec3} and compare them to empirical photometric relations and stellar evolutionary models. In Section \ref{Sec4}, we compare the methodology and results to other, similar studies, and present the conclusions in Section \ref{Sec5}.

\section{Target selection and observations} \label{Sec2}
\subsection{Target selection}
\begin{figure*}[h]
\begin{center}
    \centerline{\includegraphics[width=2\columnwidth]{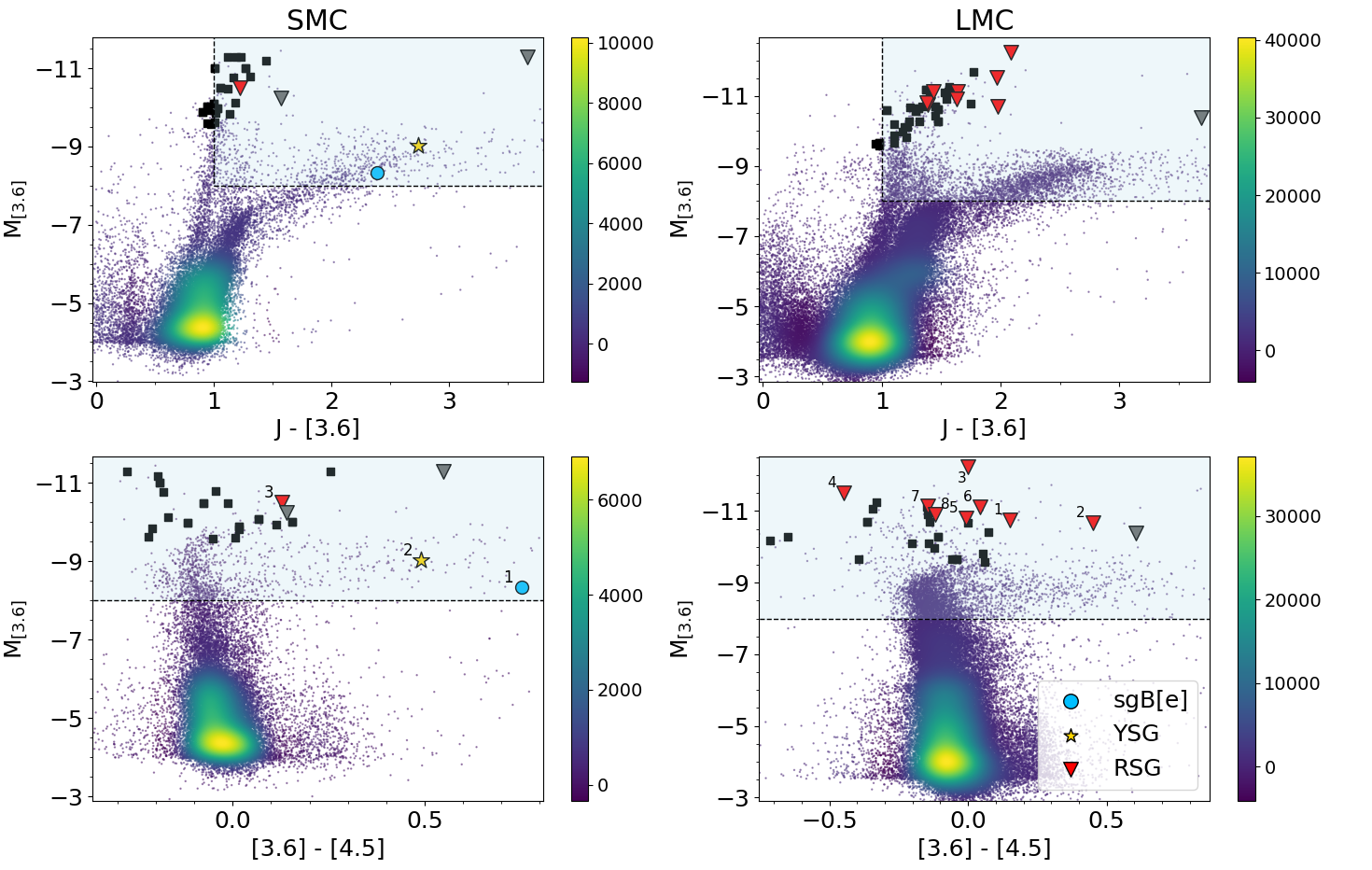}} 
    \caption{Color magnitude diagrams for the SMC (left) and LMC (right). The background points are sources from the catalogues used for target selection \citep{Yang2019a,Yang2021}. The colour bar indicates the number density of the stellar population. We show RSGs classified by \cite{Levesque2006} in black. Inverted grey triangles indicate contaminating objects that were observed (i.e. SMC4, SMC5 and LMC9). Our final sample is indicated with the coloured symbols indicated by the legend. Top: M$_{[3.6]}$ vs. $J-[3.6]$ CMD showing the criteria used to select supergiant stars. In the panel on the right, seemingly only 7 inverted red triangles are plotted, which is due to the nearly identical position of LMC1 and LMC5 in this panel. Bottom: M$_{[3.6]}$ vs. ${\rm [3.6]-[4.5]}$ CMD to visually inspect the locations of the objects using IRAC bands. The numbers correspond to the object IDs of the verified supergiants. 
    }
    \label{CMDs}
\end{center}
\end{figure*}

\begin{figure*}[h]
\begin{center}
    \centerline{\includegraphics[width=2\columnwidth]{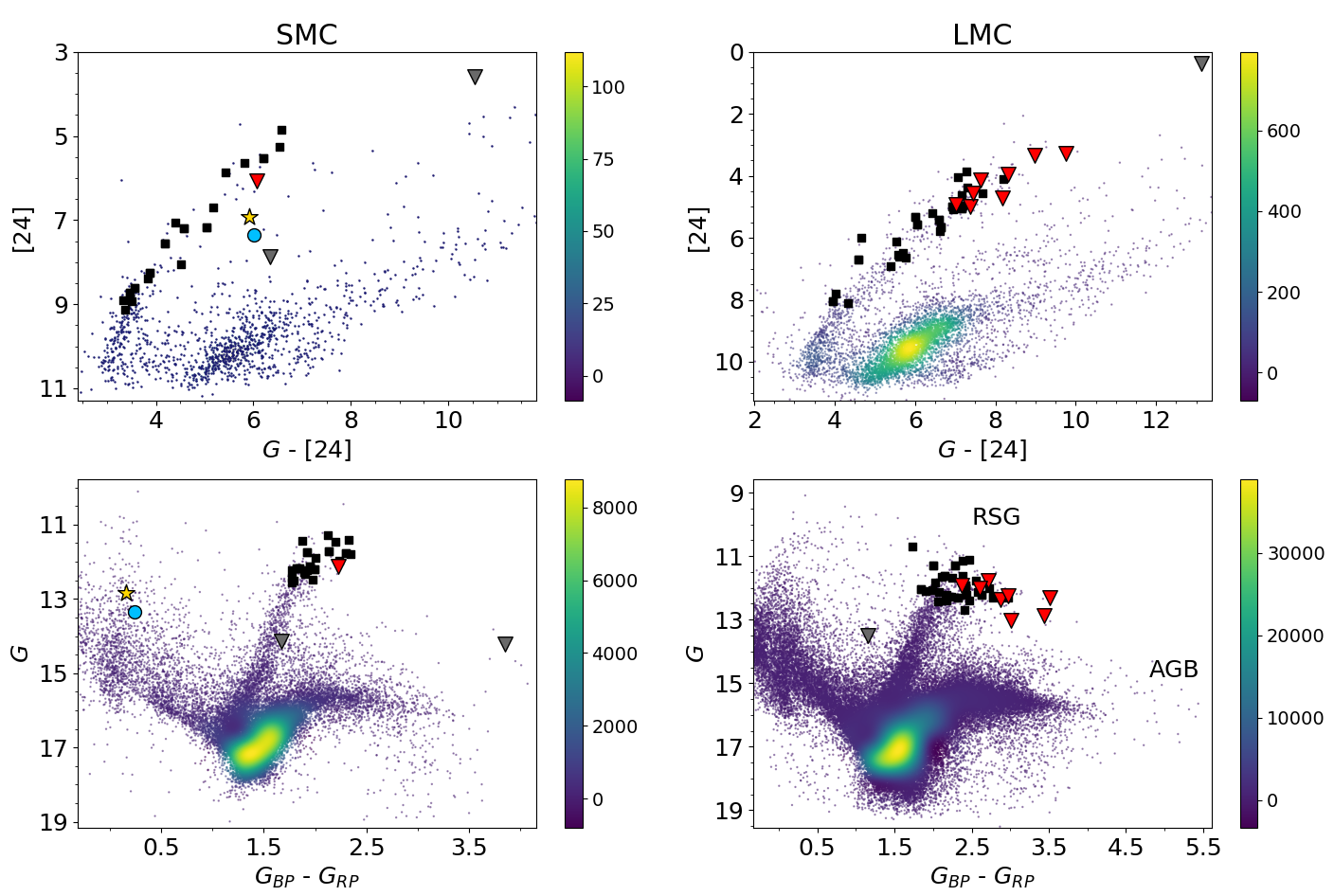}} 
    \caption{Similar to Fig. \ref{CMDs}, but for [24] vs. $\textrm G - [24]$ (top) and G vs. $\textrm G_\textrm{BP} - \textrm G_\textrm{RP}$ (bottom). The top CMD was used for the third selection criterion, while the bottom CMD was used to check whether the selected sources cluster in regions where they are expected. The RSG branch extends upwards in the middle, with the AGB branch extending to the right.}
    \label{CMDs2}
\end{center}
\end{figure*}

We based our target selection on foreground-cleaned, multi-wavelength photometric catalogues for the SMC \citep{Yang2019a} and LMC \citep{Yang2021}. These extensive catalogues comprise many photometric data sets, including \textit{Gaia} \citep{Gaia2016,GaiaDR2}, \textit{Spitzer} \citep{Werner2004} and 2MASS \citep{Cutri2003,Skrutskie2006} photometry for numerous evolved massive stars in the Magellanic Clouds. To properly select dusty supergiant candidates, we used a set of criteria to distinguish them from the asymptotic giant branch (AGB) population. Following a similar approach to \cite{Britavskiy2014}, we used the following set of criteria: \\
\\
\textit{i)}   M$_{[3.6]}$ < $-$8 mag \citep[see][]{Britavskiy2014};\\  
\textit{ii)}  J$-$[$3.6$] > 1 mag \citep[see][]{Bonanos2009,Bonanos2010};    \\
\textit{iii)} detection in [24] \\ 
\\
The above criteria allow us to select the most luminous (i.e. massive and evolved) and red sources in the near-IR and mid-IR bands. RSGs are bright sources in the near-IR due to their dramatically increased radii and low effective temperatures, while disks around  sgB[e] stars are regions in which circumstellar dust may form, making them appear bright and red in the near and mid-IR. Other potential contaminants may be Extreme-AGBs. These overlap with the region of the mid-IR CMD in which yellow supergiant stars (YSGs) and sgB[e] stars are found. \\
\indent
A detection in [$24$] indicates a cooler dusty circumstellar environment, revealing a potential preceding phase of enhanced mass loss. In Table \ref{MagTable} we list the coordinates and photometry for all selected targets. The first four columns indicate the name and coordinates of the selected targets. The following nine columns show the magnitudes used to construct the CMDs and to derive effective temperatures. \textit{Gaia} magnitudes were taken from \textit{Gaia} DR2, $V_\textrm{mean}$ from ASAS-SN, \textit{J} $\&$ \textit{K$_s$} from 2MASS, [3.6] $\&$ [4.5] from IRAC and [24] from MIPS. When available, we added time series data from the All-Sky Automated Survey for Supernovae \citep[ASAS-SN;][]{ASAS1,ASAS2}, to highlight the type and amplitude of the variability for our sources. The type (i.e. semi-regular; SR, Mira variable; M or slow-irregular; L) and amplitude of the variability are shown in the last two columns of Table \ref{MagTable}. We obtained the amplitude of the variability from the difference between the two most extreme points in the ASAS-SN light curves. The variability information improved our understanding of the sources as well as the interpretation of the uncertainties on the derived parameters. \\
\indent
Fig. \ref{CMDs} shows colour-magnitude diagrams (CMDs) in the near-IR to mid-IR for the SMC and LMC. We plot spectroscopically verified RSG populations from \cite{Levesque2006} and \cite{Davies2013} for comparison. These two works have used similar methods to derive the properties of RSGs and therefore provided the best possible comparison to our results. We then illustrate the criteria used to select evolved massive stars with the blue shaded box. The coloured markers in Fig. \ref{CMDs} make up the final sample after applying these criteria. From the top right CMD in Fig. \ref{CMDs} it is apparent that we have selected redder sources compared to other studies, implying that these sources may be dust-obscured RSGs with high extinction. Furthermore, we verified that most of our sources were clustered at the tip of the RSG branch of the optical CMD. In this CMD, AGB stars are expected to be found extending towards the right, while other contaminants are expected to be at the fainter end of the diagram (see Fig. \ref{CMDs2} bottom panel). Next, we searched the literature for spectral classifications for sources fitting the selection criteria and discarded those with accurate existing spectral classifications from the target list. The final sample of observed stars consisted of ten previously understudied objects, from which eight objects are located in the LMC and two in the SMC. Most of these objects had been assigned a spectral type in the past, but no luminosity class, implying they have not been previously (spectroscopically) verified as RSGs. One object was added to the sample (SMC1) due to its rare class (sgB[e]; e.g. \citealt{Kraus2019}), to investigate evidence of variations in its spectral type, completing the sample to eleven stars. We also retained three sources that partially satisfied our selection criteria, to investigate potential contaminants; i.e.\ altogether we observed fourteen sources. \\

\begin{table*}
\centering
  \begin{threeparttable}
    \caption{Photometric data of the sample.}
    \tiny
    \label{MagTable}
        \begin{tabular}{c c c c c r r r r r r r r l l}        
        \hline\hline                 
        ID & SIMBAD Name & R.A. & DEC & $G$ & $G_\textrm{BP}$ & $G_\textrm{RP}$ & $V_\textrm{mean}$ & $J$ & $K_S$ & $[3.6]$ & $[4.5]$ & $[24]$ & Var. & $V_\textrm{amp}$\\    
         &  & (J2000) & (J2000) & (mag) & (mag)& (mag)& (mag)& (mag)& (mag)& (mag)& (mag)& (mag) & Type & (mag) \\  
        \hline\hline                        
            SMC1 & LHA 115-S 23 & 00 55 53.81 & -72 08 59.51 & 13.32 & 13.39 & 13.16 & 13.36 & 13.01 & 12.18 & 10.63 & 9.87 & 7.34 & - & -\\
            SMC2 & $[$MA93$]$ 1810 &  01 15 21.22 & -73 30 15.10 & 12.82 & 12.87 & 12.71 & 12.80 & 12.67 & 11.69 & 9.94 & 9.45 & 6.92 & - & - \\
            SMC3 & PMMR 64 &  00 53 53.08 & -71 42 44.37 & 12.10 & 13.29 & 11.06 & 12.89 & 9.67 & 8.65 & 8.46 & 8.33 & 6.06 & L & 1.50 \\
            SMC4 & IRAS 00350-7436 & 00 36 59.56 & -74 19 50.26 & 14.17 & 14.92 & 13.25 & 14.56 & 11.32 & 9.13 & 7.67 & 7.12 & 3.58 & - & 0.19\\
            SMC5 & SV* HV 859 & 01 10 26.92	& -72 35 48.57 & 14.18 & 16.52 & 12.66 & 15.26 & 10.28 & 9.08 & 8.72 & 8.58 & 7.86 & M & 3.80 \\
         \hline    
            LMC1 & WOH S 57 &  04 53 14.80	& -69 12 18.00  & 12.34 & 13.98 & 11.11 & 13.59 & 9.14 & 7.75 & 7.77 & 7.62 & 4.99 & SR & 1.14\\
            LMC2 & WOH S 374 &  05 31 47.42 &	-66 03 40.58  & 13.01 & 14.78 & 11.78 & 14.40$^1$ & 9.80 & 8.67 & 7.84 & 7.39 & 3.27 & Var:$^1$ & -\\
            LMC3 & $[$W60$]$ B90 &  05 24 19.31 &	-69 38 49.37  & 12.29 & 14.47 & 10.95 & 15.11 & 8.36 & 6.81 & 6.29 & 6.29 & 3.35 & L & 0.38\\
            LMC4 & UCAC2 2674864  &  05 44 13.77 &	-66 16 44.62 & 12.86 & 14.99 & 11.54 & 14.28 & 8.96 & 7.49 & 7.01 & 7.46 & 4.71 & SR$^2$ & 0.31$^3$ \\
            LMC5 & SP77 28-2 &  04 57 26.36 & -66 23 25.78 & 11.91 & 12.99 & 10.61 & 13.03	& 9.09	& 7.88 & 7.72 & 7.73 & 4.92 & SR & 1.54\\
            LMC6 & SV* HV 12185 &  05 09 43.58	& -65 21 59.18 & 11.74 & 13.30 & 10.59 & 12.97 & 8.83 & 7.69 & 7.41	& 7.36	& 4.12 & SR & 1.49\\
            LMC7 & SV* HV 12793 &  05 23 43.61  & -65 41 59.84  & 12.23 & 13.98 & 11.01 & 13.57 & 9.01 & 7.74 & 7.39 & 7.53 & 3.94 & SR & 1.11 \\
            LMC8 & W61 19-24 &  05 32 20.19 &	-67 32 42.36 & 12.00 & 13.45 & 10.85 & 12.30 & 9.22 & 8.01 & 7.60 & 7.72 & 4.57 & SR & 0.71\\
            LMC9 & MSX LMC 806 & 05 32 03.42 & -67 42 25.70 & 13.47 & 13.91 & 12.75 & 13.69 & 11.82 & 10.04 & 8.15 & 7.54 & 0.37 & - & - \\
        \hline
        \end{tabular}
    \begin{tablenotes}
        \small
        \vspace{5pt}
        \item \textit{Notes:} The variability type indicates the following classes: semi regular (SR), mira variable (M) and slow irregular (L) variables. \\$^1$Magnitude and variability designation as listed by \cite{Samus2017}. \\$^2$Designation from \textit{Gaia} DR2. \\$^3$Semi-amplitude from \cite{Groenewegen2018}.
    \end{tablenotes}
    
  \end{threeparttable}
\end{table*}

\subsection{Spectroscopic observations}
We obtained optical spectra for the fourteen selected targets with the 6.5m Baade telescope at Las Campanas Observatory, Chile. The targets were observed with the Magellan Echellette (MagE) spectrograph \citep{Marshall2008a} using the 1" science slit. MagE is a medium resolution (resolving power $\rm {R} \sim 4100$ for the 1" slit) single-object spectrograph and provides spectral coverage in the entire optical domain ($\lambda$~$3200 - 10000\,\AA$). Multiple exposures were taken for each object. The observations were completed in four separate observing runs between December 2018 and March 2020. A log of the observations is presented in the first three columns of Table \ref{ObsTable}. \\

\begin{table*}[h]
\centering     
  \begin{threeparttable}
    \caption{Observational properties.}
    \tiny
    \label{ObsTable}
        \begin{tabular}{l l l l l l l l} 
        \hline\hline                 
        ID & Obs. Date  & Exp. Time & S/N & RV & New Spectral & Previous Spectral & References\\
         & (UT) & (s) &  & (km s$^{-1}$) & Type & Type & \\ 
        \hline\hline                        
            SMC1 & 19-Dec-2018 & 3x200 & 80 & 155 $\pm$ 4 & A0 I[e] & B8~I[e], A1~I[e]&\cite{Zickgraf1992}, \cite{Kraus2008}\\
            SMC2 & 19-Dec-2018 & 3x180 & 85 & 189 $\pm$ 10 & F8 I & ... & ...\\
            SMC3 & 13-Nov-2019 & 3x100 & 95 & 178 $\pm$ 7 & M2 I & K/M &\cite{SanduleakSMC3}\\
            SMC4 & 12-Sep-2019 & 3x600 & 65 & ... & C0-2,2e & C3,2e &\cite{Groenewegen1998}\\
            SMC5 & 13-Nov-2019 & 3x300 & 20 & ... & M7e III & K5/M7e & \cite{Wood1983}\\
            \hline  
            LMC1   & 12-Nov-2019 & 3x200  & 60 & 272 $\pm$ 8 & M2 I & M0:&\cite{Westerlund1981}\\
            LMC2  & 13-Nov-2019 & 3x150  & 45 & 306 $\pm$ 11 & M4-5 II-III & M6&\cite{LMC2class}\\
            LMC3  & 08-Mar-2020 & 3x180,1x240  & 55 & 260 $\pm$ 6 & M3 I & M2&\cite{LMC3class}\\
            LMC4  & 08-Mar-2020 & 3x180  & 50 & 296 $\pm$ 6 & M0 I & M0.5&\cite{LMC3class}\\
            LMC5 &  08-Mar-2020 & 3x100  & 55 & 290 $\pm$ 5 & M1 I & M0&\cite{Westerlund1981}\\
            LMC6  & 08-Mar-2020 & 3x100  & 75 & 318 $\pm$ 5 & M1 I & M1 I:&\cite{Samus2017}\\
            LMC7  &  08-Mar-2020 & 3x180  & 70 & 316 $\pm$ 9 & M2 I & M3/M4&\cite{LMC2class}\\
            LMC8  & 08-Mar-2020 & 3x120,1x100 & 65 & 304 $\pm$ 5 & M3 I & M2&\cite{LMC8class}\\
            LMC9  & 08-Mar-2020 & 3x180 & 70 & ... & H~\textsc{ii} + LBVc & YSO: &\cite{LMC9class}\\

        \hline
        \end{tabular}
    \begin{tablenotes}
        \small
        \vspace{5pt}
        \item \textit{Notes}: The signal-to-noise ratio (S/N) of the combined spectrum is measured in the red ($\sim$8500$\AA$). No previous luminosity classes were indicated for most stars. The previous classification of LMC6 included a luminosity class, but this was an uncertain classification.
    \end{tablenotes}
    
  \end{threeparttable}
\end{table*}

\subsection{Data reduction}
The spectroscopic data were reduced using the MagE Spectral Extractor \citep[MASE;][]{Bochanski2009a}. The MASE pipeline offers the full reduction process from raw data to a 1D spectrum, including the bias subtraction, flat fielding, sky subtraction, cosmic ray removal, flux and wavelength calibration and finally the extraction of the echelle orders to 1D-spectra for all exposures. After carefully transforming the wavelength units from vacuum to air wavelengths, which is required for spectral modelling, we merged the individual exposures using the median flux in every wavelength bin. The 1D echelle orders were then connected using linear weighting at overlapping wavelengths to assemble the full 1D spectrum. \\
\indent
Artificial bumps were present in the merged spectrum, caused by the small offset between orders. We have corrected for this by scaling down the individual orders so that their relative fluxes matched at overlapping wavelength sections. Once the spectrum was cleaned for these bumps, we scaled the flux back up in units of absolute flux. The spectra were not corrected for tellurics. The S/N presented in the fourth column of Table \ref{ObsTable} has been derived by calculating the median S/N in small chunks of pseudo-continuum in the \textit{I}-band and serves as a general indicator of the spectral quality (S/N $\geq 50$ in most cases).\\

\subsection{RV measurements}

Before proceeding with the spectral modelling, the spectra were corrected for the radial velocity (RV) shift. Specifically, the Ca~\textsc{ii} triplet ($\lambda$8498, $\lambda$8542, $\lambda$8662), the Na~\textsc{i} doublet ($\lambda$5890, $\lambda$5896) and two resolved single metal lines close to the Ca~\textsc{ii} triplet (Ti~\textsc{i} $\lambda$8426 and Fe~\textsc{i} $\lambda$8612) were used to estimate the RV\footnote{See table C.1 in \citet{Dorda2016} for a complete atlas of spectral lines for cool supergiants in the $i$ and $z$-bands.}. We measured the central wavelength for the selected metal lines by fitting a Gaussian to the line profile. Then, we calculated the offset from the rest-wavelength for these lines to determine a velocity shift for each line profile. Precisely determined values for the rest-wavelength of spectral lines were taken from the NIST atomic database \citep{NIST}. The final RV was determined as the weighted mean of the shifts of individual lines. We then applied this RV shift to the spectrum. Towards mid-late M-types, prominent metal lines disappear due to the increasing strength of TiO $\lambda$8432-8442-8452. In our sample this only applied to LMC2, for which we have derived the velocity shift solely from the Ca~\textsc{ii} triplet. The mean radial velocity for each target is presented in the sixth column of Table \ref{ObsTable}. We note that the radial velocities we have found are consistent with the systemic velocity of the LMC and SMC \citep[260~$\rm km \, s^{-1}$ and 150~$\rm km \, s^{-1}$, respectively,][]{McConn2012, Marel2014} confirming their membership to these systems. When a RV value from \textit{Gaia} DR2 was available, we compared our results and confirmed them to be consistent. Additionally, the improved \textit{Gaia} EDR3 parallax \citep{GaiaEDR3} was inspected.  We found the parallaxes of all objects, considering their errors and the systematic correction from \cite{LindegrenGaiaEDR3} to be consistent with zero, which is in turn consistent with the extragalactic origin of the sources.

\subsection{Spectral classification} \label{Sec2.5}
\subsubsection{Classification of RSGs} \label{Sec2.5.1}
\begin{figure*}
\begin{center}
    \centerline{\includegraphics[width=2.2\columnwidth]{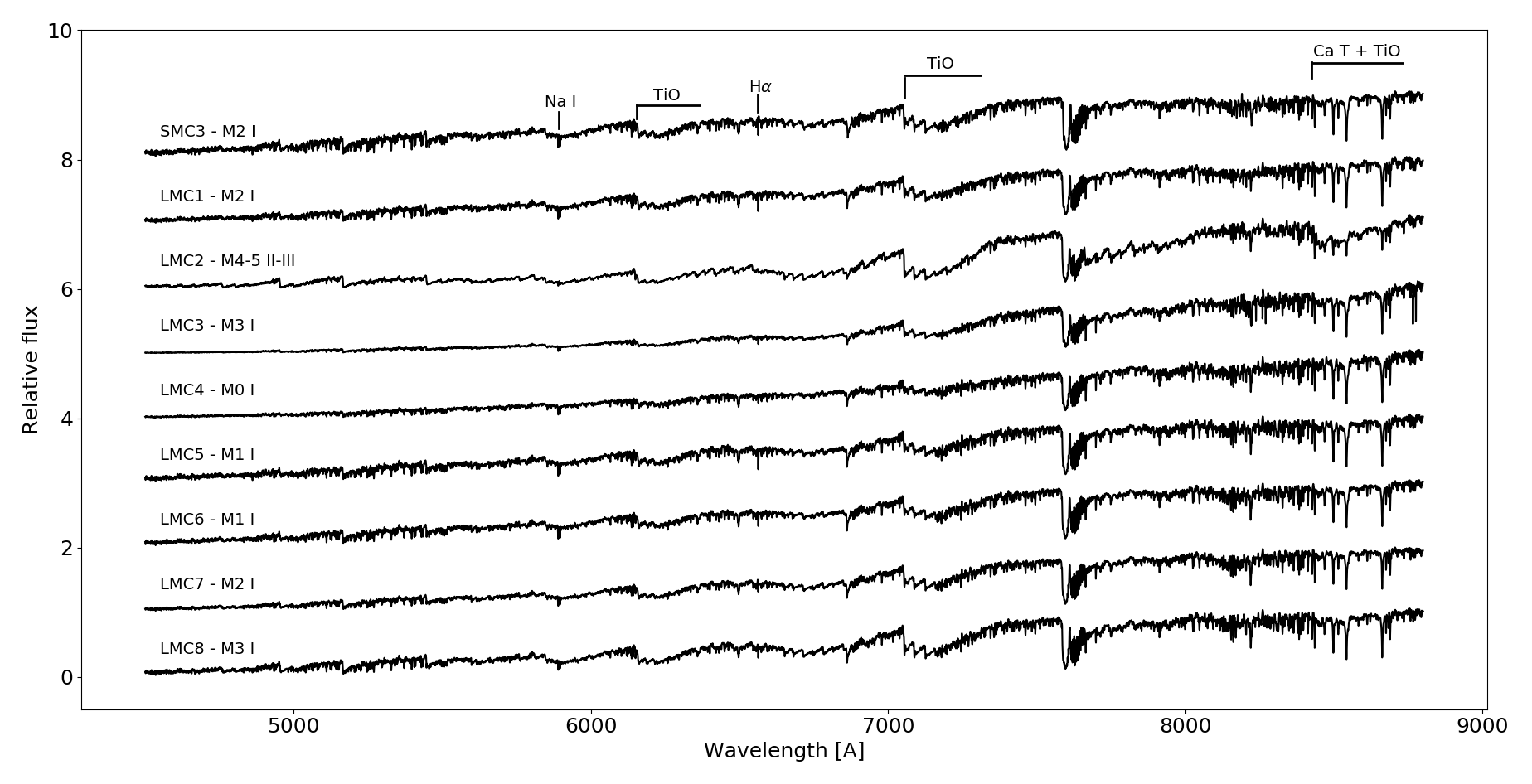}} 
    \caption{Spectra from M-type sources in our sample (an arbitrary offset has been applied for illustration purposes). The Ca~\textsc{ii} triplet and key TiO bands are indicated.}
    \label{SpectraZoomOut}
\end{center}
\end{figure*}

\begin{figure*}
\begin{center}
    \centerline{\includegraphics[width=2.3\columnwidth]{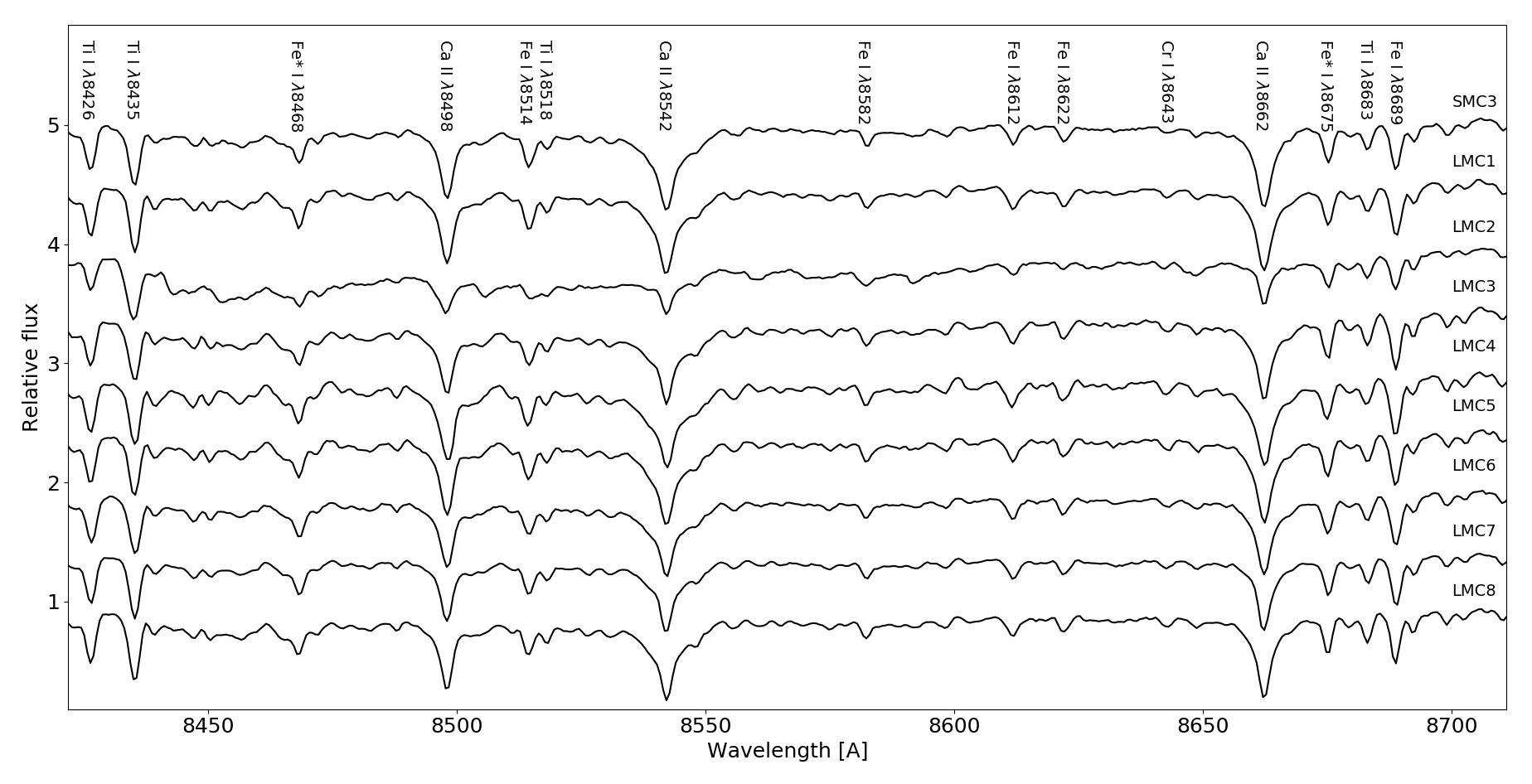}} 
    \caption{A zoom-in on the Ca~\textsc{ii} triplet and prominent metal lines \citep[see][]{Dorda2016} used for the classification of the objects and RV analysis. Features marked with a * are blends and only the most significant line is indicated. For LMC2, Ti~\textsc{i} $\lambda$8435 is relatively strong and the ratio between Fe~\textsc{i} $\lambda$8514 and Ti~\textsc{i} $\lambda$8518 line strengths is close to unity.}
    \label{SpectraZoomIn}
\end{center}
\end{figure*}

The abundant presence of molecular bands in the optical are a clear indicator of late-type stars. Depending on the metal content of the star, the relative strengths of the molecular bands are typically used to determine the spectral sub-type of M-stars. Despite being derived from strong and abundant molecular bands, the optical classification of M0-M3 stars posed difficulties, as the relative strengths of the TiO bands change only slightly in this range, while metal lines at $\sim$8400-8800$\AA$ remain unchanged due to their insensitivity to temperature \citep{Solf1978}. For later types (M4-M6), \cite{Solf1978} indicated the blending of the Ti~\textsc{ii} $\lambda$8435 doublet with the triple TiO band-head at $\lambda$8432-8442-8452 as the primary classifier.\\
\indent
Given these spectral classification criteria, we then proceeded with the classification of our spectra by visual inspection. Nine stars were identified as M-type stars due to the presence of strong molecular TiO bands at $\lambda$6150 and $\lambda$7050. Fig. \ref{SpectraZoomOut} presents the optical spectra of these stars. The most important spectral features are indicated and a spectral type was added. For seven of these stars, we identified similar TiO band strengths amongst them. For this group, we sorted the spectra from weakest TiO $\lambda$7050 band (M1) to strongest (M3). Intermediate objects were classified as M2. The remaining two stars (LMC2 and LMC4) were noticeably different from the rest of the sample. Fig. \ref{SpectraZoomIn} highlights spectral lines in the Ca~\textsc{ii} triplet region, for example Ti~\textsc{ii} $\lambda$8435. For LMC2, the TiO $\lambda$8432 band is present and blends with Ti~\textsc{ii} $\lambda$8435, but does not dominate the metal lines in this region. When the spectrum is depleted of metal line features due to the increasing TiO $\lambda$8432 band strength, the star is considered as spectral type M6 \citep{Negueruela2012}. Therefore, we classified LMC2 as M4-5. For LMC4, the TiO bands at $\lambda$6150 and $\lambda$7050 were present, but were noticeably weaker compared to the other stars, indicating an early M type. We classified this object as M0. \\
\indent
For the luminosity class we used the strength of the Ca~\textsc{ii} triplet ($\lambda$8498, $\lambda$8542 $\&$ $\lambda$8662) as the main criterion. The Ca~\textsc{ii} triplet is sensitive to changes in surface gravity due to pressure broadening, but insensitive to $T_{\rm eff}$ and $Z$ changes \citep{Massey1998a}. Therefore, a strong Ca~\textsc{ii} triplet is expected for supergiant stars. A strong Ca~\textsc{ii} triplet was indeed detected in all stars, with the exception of LMC2 (see Fig. \ref{SpectraZoomIn}). Apart from the Ca~\textsc{ii} triplet, the ratio between Fe~\textsc{i} $\lambda$8514 and Ti~\textsc{i} $\lambda$8518 was indicative (except LMC2, where neither line clearly dominates the other) of luminosity $\textrm{class I}$  \citep{Britavskiy2014}. LMC2 was classified as M4-5~II-III, as it does not satisfy the criteria to be classified as a luminosity $\textrm{class I}$ star. To our knowledge, criteria for luminosity class II are not well established. Therefore, we cannot reject this possibility and present a range of luminosity class II-III for this object. \\
\indent
The variability types presented in Table \ref{MagTable} support the spectral classification, as stars of type semi-regular (SR) and slow-irregular (L) are often attributed to the RSG or red-giant class. Furthermore, most of our classified RSGs show minimum to modest (up to 1.5 magnitudes) variability in the optical, which is expected for RSGs. The final classification of our objects, along with a comparison to older classifications and references for these, is presented in the final columns of Table \ref{ObsTable}.\\

\subsubsection{Classification of other supergiants}
During the observing campaign, two additional evolved massive stars were observed. We present spectra for these in Fig. \ref{YSGBSG}, \ref{LHA115S23Z1}, \ref{LHA115S23Z2} and \ref{MA93Z1}. The nature of their spectra was such that they could be classified, but needed higher spectral resolution for adequate modelling (i.e. the models constrain properties from narrow metal line ratios), and this was not pursued. We briefly describe their spectral types here.\\
\indent
The spectrum of SMC1 displayed strong double peaked emission for the Balmer series and several forbidden [O~\textsc{i}] and [Fe~\textsc{ii}] emission lines (see Fig. \ref{LHA115S23Z1}). These features originate from a circumstellar disk or ring structure \citep{Maravelias2018}. Despite being previously classified as a B8~I[e] \citep[][spectrum taken in December 1989]{Zickgraf1992} and as A1~I[e] \citep[][spectrum taken in October 2000]{Kraus2008}, we determined a spectral type of A0~I[e] (spectrum taken in December 2018) from the Mg~\textsc{ii} $\lambda$4481 to He~\textsc{i} $\lambda$4471 line ratio (see Fig. \ref{LHA115S23Z2}). If Mg~\textsc{ii} $\lambda 4481 \geq$ He~\textsc{i} $\lambda$4471, the spectral type is later than B9, while when the ratio of Ca~K/(H$\epsilon + $Ca~H)~$\leq 0.33$, the spectral type is A0 \citep{Evans2004}, which is the case for our spectrum. We also note that the He~\textsc{i} lines reappeared in our spectrum, after being absent in the spectrum analysed by \cite{Kraus2008}, making the spectrum appear slightly hotter than it was in \cite{Kraus2008}. Assuming the spectra have been classified correctly in past studies, the spectral type of SMC1 apparently varied on the scale of decades. For the radial velocity measurement, several metal lines in the optical wavelength range were used, such as Mg~\textsc{ii} $\lambda$4481, He~\textsc{i} $\lambda$4471, 5875, Si~\textsc{ii} $\lambda$6347 and Fe~\textsc{ii} $\lambda$4174, 4179. Modelling of this source was not pursued due to the complexity of modelling the disk emission features and their contaminating effect on key Balmer and He~\textsc{i} lines.\\
\indent
SMC2 is a yellow supergiant star that displayed a strong H$\alpha$ emission component, indicating a mass-loss component surrounding the star. The spectrum is characterized by a series of Ti and Fe lines in the $4000-5000\,\textrm{\AA}$ region and strong Ca~\textsc{ii} H and K absorption (see Fig. \ref{MA93Z1}). We classified this star as F8~I, due to the absence of the G-band (< G0) and the strong Ti-Fe forest lines relative to the Balmer lines at $Z = 0.2\,Z_{\odot}$ (> F5). The sharpness of the Balmer lines and the strengths of the metal lines suggests that this star is of luminosity class I. For the radial velocity, several resolved metal lines were used, such as Mg~\textsc{ii} $\lambda$4481, Si~\textsc{ii} $\lambda$6347 and Na~\textsc{i} $\lambda$5890, 5896. For yellow supergiants, spectral models are scarce. Models for a $T_{\rm eff}$ estimate based on Fe line ratios have recently been published \citep{Kourniotis2022}, however they require us to resolve narrow Fe lines. Our current spectrum is not able to resolve these lines and therefore a higher resolution spectrum is needed to be able to derive a $T_{\rm eff}$ empirically.

\begin{figure*}
\begin{center}
    \centerline{\includegraphics[width=2.1\columnwidth]{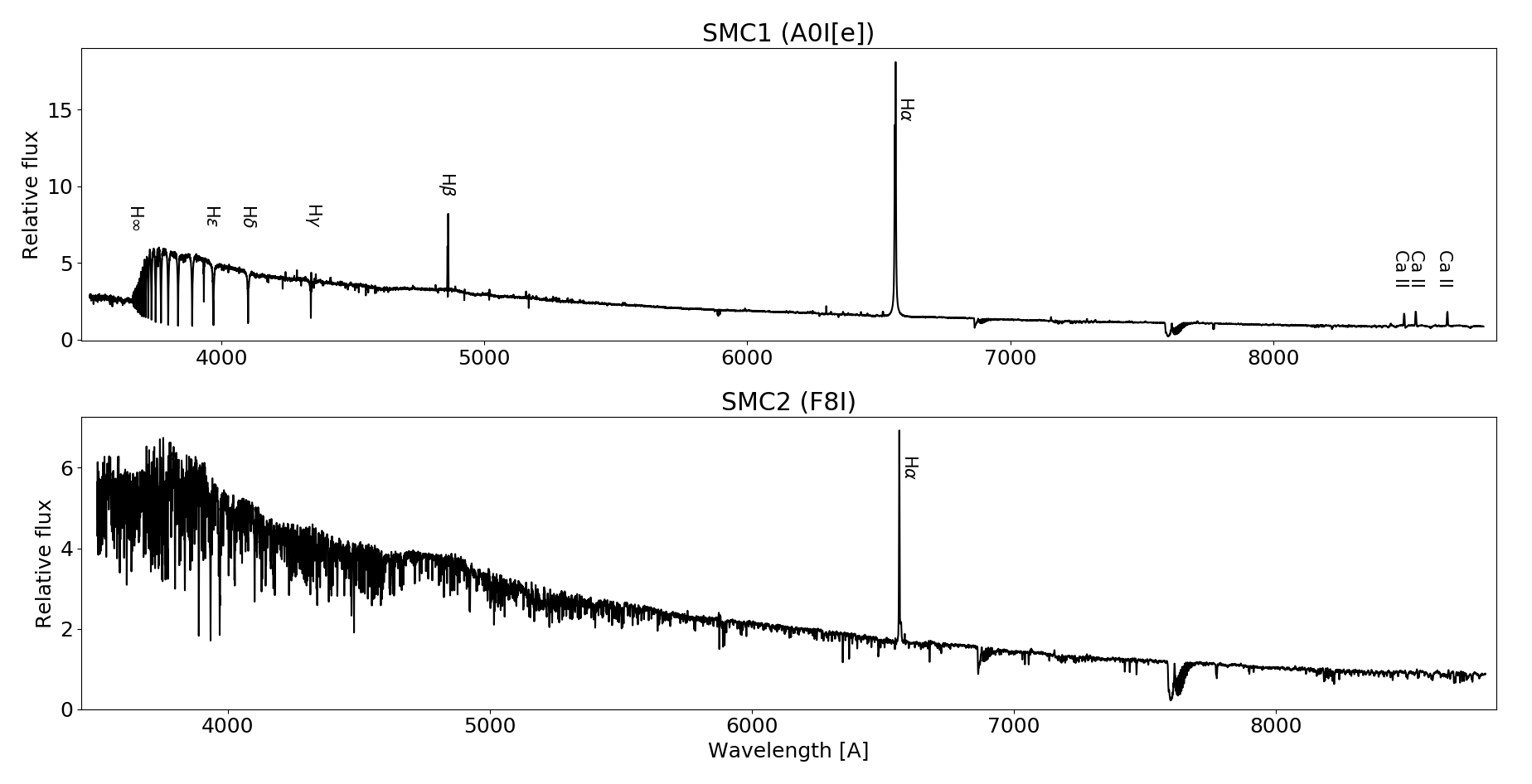}} 
    \caption{\textit{Top}: Spectrum of SMC1, an A0~I[e] star. The spectrum is characterized by several strong emission lines, which are indicated. \textit{Bottom}: Spectrum of SMC2, a F8I star. Strong H$\alpha$ emission was detected indicating expanding circumstellar material.}
    \label{YSGBSG}
\end{center}
\end{figure*}

\subsubsection{Classification of other sources}
Three more objects fitting the selection criteria were observed (see Fig. \ref{CMDs} and \ref{CMDs2}, inverted grey triangles), but were not classified as evolved massive stars. We found SMC4 and SMC5 to have characteristic features of giant stars, while LMC9 shows the characteristics of an ionized nebula. \\
\indent
The spectrum of SMC4 contains carbon absorption bands at $\lambda$4737 and $\lambda$5165 and deep P-Cygni profiles indicative of strong mass loss \citep{Whitelock1989}, which could explain its extreme magnitude at [24]. SMC4 is a spectroscopic binary, with pairs of shifted spectral lines. We classified the primary star as a carbon star with moderately strong Swan C$_2$ bands at $\lambda$4737 and $\lambda$5165, but with the absence of the C$_2$ $\lambda$5585 band. Due to the strength of the Balmer lines, the absence of Ca~\textsc{i} $\lambda$4426 and the presence of the G-band, albeit weak compared to H$\gamma$, we conclude this star is of G-type, and its carbon star equivalent should therefore be in the range C0-2. We were not able to obtain a more secure classification due to the contaminating features of the companion star. \\
\indent
SMC5 is characterized by extreme TiO absorption bands. The relative strengths of the TiO bands are indicative of a late M-star, beyond M6. The strength of VO $\lambda$7900 suggests a spectral type of M7. The strength of the Ca~\textsc{ii} triplet indicates the star is of luminosity class III. \\
\indent
The spectrum of LMC9 was characterized by strong, narrow emission lines, most notably the [O~\textsc{iii}] feature at $\lambda$5007 and a saturated H$\alpha$ emission component. Most of the emission lines were identified as recombination lines from low ionisation states, typically present in a H\textsc{ii} region. A stellar continuum has been observed underneath the sharp emission lines, revealing a bright blue object. From the broadened Balmer line profiles and Fe~\textsc{ii} emission lines present throughout the spectrum we conclude that a Luminous Blue Variable candidate (LBVc) may be present inside this H\textsc{ii} region, although further investigation is needed to verify this object as such.

\begin{figure*}
\begin{center}
    \centerline{\includegraphics[width=2.1\columnwidth]{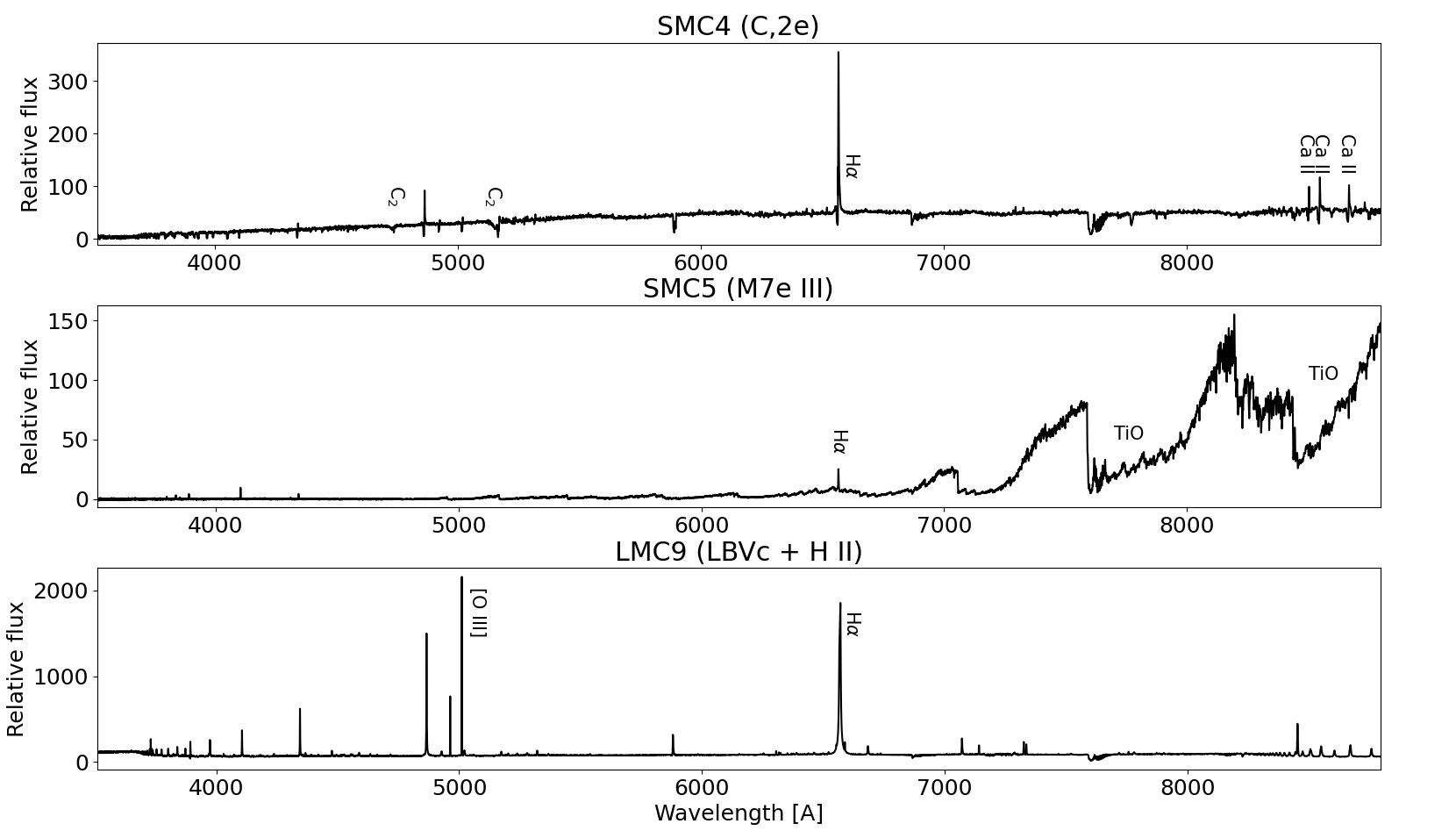}} 
    \caption{\textit{Top}: Spectrum of SMC4, a carbon star. The spectrum is characterized by two molecular carbon bands and several strong, deep P-Cygni profiles. \textit{Middle}: Spectrum of SMC5, a late M giant star. Strong TiO bands are revealed in the red part of the spectrum, indicating a low effective temperature. From the Ca~\textsc{ii}triplet, only $\lambda$8662 is present, albeit almost completely blended in with overarching TiO band. \textit{Bottom}: Spectrum of LMC9, an ionized nebula. The overall spectrum is dominated by strong emission lines coming from a nebula. Strong [O~\textsc{iii}] was detected and the H$\alpha$ emission was saturated. The emission line profile of H$\alpha$ shows significant broadening and other circumstellar lines (Fe~\textsc{ii}) were detected in the left wing of H$\alpha$, revealing a LBVc.}
    \label{YSGBSG}
\end{center}
\end{figure*}

\section{Spectral modelling and resulting parameters} \label{Sec3}
\subsection{Fitting the \textsc{marcs} models} \label{Sec3.1}
\begin{figure*}
\begin{center}
    \centerline{\includegraphics[width=2.2\columnwidth]{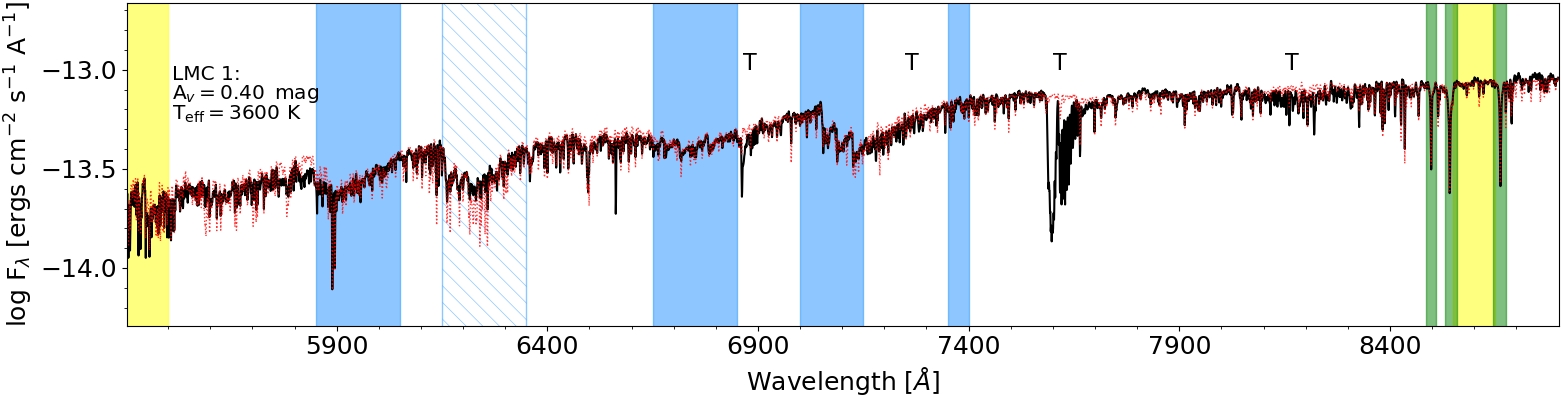}} 
    \caption{Best fit \textsc{marcs} model (red dotted line) to the spectrum of LMC1 (black solid line). Regions included in the $\chi^2_ \textrm {red,min}$ calculation are indicated with shaded areas. The best fit properties are indicated in the top left corner. Tellurics are indicated with a T, the Ca \textsc{ii} triplet with green shades, the regions used for $A_V$ in yellow shades and the TiO bands are indicated with blue shades. The TiO band at $\lambda$6150 (hatched) was excluded from the fit.}
    \label{BestFit16}
\end{center}
\end{figure*}

\begin{figure}
\begin{center}
    \centerline{\includegraphics[width=1.05\columnwidth]{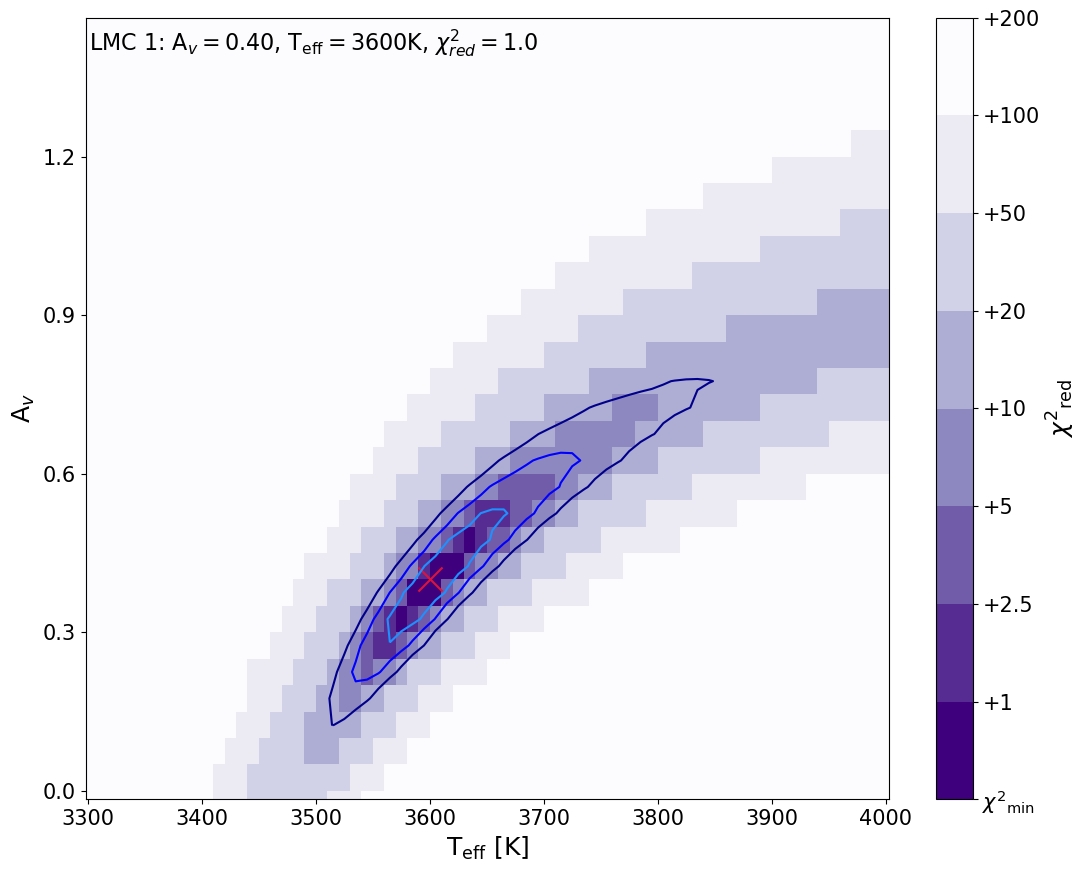}} 
    \caption{A map of $A_V$ and $T_{\rm eff}$ for LMC1, with darker shades indicating parts of the grid with lower $\chi^2$ values. The $\chi^2$ values of the color bar are with respect to the minimum $\chi^2$. Contours at 1$\sigma$, 2$\sigma$ and 3$\sigma$ are indicated to estimate the uncertainties.}
    \label{Chisq16}
\end{center}
\end{figure}

We used a grid of alpha-poor, spherical \textsc{marcs} model atmospheres \citep{Gustafsson2008} to fit the M-type stars in our sample. The computed models have a mass of $15\,\textrm{M}_{\odot}$. Given the expected mass range for stars that evolve towards the RSG phase (8M$_{\odot}$ $\leq$ M$_*$ $\leq$ 25M$_{\odot}$) and the availability of models with limited discrete stellar masses ($0.5, 1, 2, 5\, \& \,15\,\textrm{M}_{\odot}$), the 15$\,\textrm{M}_{\odot}$ model was the most suitable for our studies. Furthermore, we assume that a single mass to represent the entire range of masses for RSGs is justified given that the geometrical thickness, and thus the atmospheric structure is largely unaffected in this mass range \citep{Davies2010}. The microturbulent velocity of the models was fixed to $\xi=5\,\textrm{km s}^{-1}$.
 \cite{Gonzalez2021} have shown that changes in $\xi$ had little effect on the final result. We used \textsc{marcs} models with log$(Z/Z_{\odot})=-0.35\,\textrm{dex}$ for the LMC and log$(Z/Z_{\odot})=-0.55\,\textrm{dex}$ for the SMC, which adequately represent the average metallicity derived empirically from a set of RSGs in the Magellanic Clouds \citep{Davies2015}. The surface gravities vary between -0.5 and +$0.5\,\textrm{dex}$ in steps of $0.1\,\textrm{dex}$. Only models with a surface gravity of +0.5, +0.0 and -$0.5\,\textrm{dex}$ were available from the \textsc{marcs} platform, hence we interpolated the flux of adjacent models linearly to make the grid more dense. The effective temperature ranges from $3300-4500\,\textrm{K}$, with $10\,\textrm{K}$ steps for the range of $3300-4000\,\textrm{K}$ and $25\,\textrm{K}$ steps for the range of $4000-4500\,\textrm{K}$ using similar interpolation strategies.\\
\indent
We first degraded the resolution of the models from $R=20000$ to $R = 4000$, so that the model spectrum matches the spectral resolution of MagE. We then applied \citet{Fitzpatrick1999} reddening laws with varying $A_V$ to the models to derive the best fit extinction factor, assuming a typical total-to-selective extinction of $R_V=2.74$ and $R_V=3.41$ for the SMC and LMC, respectively \citep{Gordon2003}. We note that the values for $R_V$ may be higher, considering the grain size distribution in the circumstellar environment of RSGs \citep{Massey2005}. A recent study by \cite{Gonzalez2021} however, have indicated that changes in $R_V$ and the type of extinction law used, do not largely contribute to temperature changes of the best fit model. To obtain the best fit model to our spectra, we computed the reduced chi-squared ($\chi^2_{\rm red}$):

\begin{ceqn}
\begin{align}
    \chi^2_\textrm{red} = \frac{1}{\nu} \sum^\nu_i \left( \frac{F_i - F_{i,\textrm {mod}}}{\sigma_i}\right)^2,
    \label{eq:chisq}
\end{align}
\end{ceqn}

\noindent
where $\nu$ is the degrees of freedom, $F_i$ and $F_{i, \textrm {mod}}$ are the fluxes of the spectrum and model, respectively, and $\sigma_i$ is the uncertainty on the flux in bin $i$. The degrees of freedom are set by the amount of bins minus the amount of free parameters ($T_{\rm eff}$ and $A_V$). The accepted best-fit model was the model on the 2D grid with lowest $\chi^2_{\rm red}$ (i.e. $\chi^2_{\rm min}$). As in \citet{Gonzalez2021}, we chose to smooth the spectra to determine the $\chi^2_{\rm red}$, given that there may be uncertainties in the molecular transitions in the \textsc{marcs} models. Bin by bin fitting of the single wavelength bins may therefore yield unrealistic $\chi^2_i$, which could heavily impact the $\chi^2_\textrm{red}$. For both the model and the observed spectrum, we grouped smaller wavelength bins into larger bins of $50\,\textrm{\AA}$ and set the mean flux of the new bin as the corresponding data point. We then calculated the $\chi^2_i$ of the larger bin using the mean flux and mean error (i.e. the standard deviation of the flux measurements). We fitted selected wavelength bins between $5400\,\textrm{\AA}$ and $8800\,\textrm{\AA}$, which include the most prominent and temperature sensitive TiO bands as well as the bluest and reddest $100\,\textrm{\AA}$ to describe the slope of the spectrum to estimate $A_V$. Upon inspecting the $\chi^2_{\rm red}$ values after the initial run, we chose to discard the TiO band at $6150\,\textrm{\AA}$ from the calculation, as the molecular transitions in the models do not accurately match the features of real RSG spectra (see Fig. \ref{BestFit16}) and therefore skew the results. As our spectra were not corrected for telluric contamination, we avoided bins including telluric bands. Isolated telluric lines may still be present \citep{Catanzaro}, but given the use of the aforementioned smoothing technique, this does not significantly affect the $\chi^2_{\rm red}$ calculation. \\
\indent
We approached the modelling as follows: we first fit the Ca \textsc{ii} triplet to derive the surface gravity (log $g$). The Ca \textsc{ii} triplet is the primary diagnostic sensitive to the pressure broadening in the optical, so fitting this narrow spectral domain allowed us to get a precise measurement of the surface gravity. We proceeded by fixing the derived value of log $g$ for the remainder of the fitting routine. We then fitted a 2D grid of models with variable $A_V$ and $T_{\rm eff}$ to the spectrum and located the point on the grid with $\chi^2_{\rm min}$. An example fit is presented in Fig. \ref{BestFit16}, in which we show the observed spectrum in black and the model in red (see Fig. \ref{BestFitAll} in the Appendix for the remainder of objects). From here on, we combine the extracted best fit model parameters with photometry to derive the luminosity and subsequent radius of the star. For this, we used the $K-$band magnitude ($m_{K}$), an appropriate bolometric correction for the $K-$band based on the spectral type \citep{Davies2018} and the distance modulus ($\mu_{\rm LMC}$ $=18.477\pm0.030$ and $\mu_{\rm SMC}$ $=18.95\pm0.07$ mag; \citealt{Pietr2019,Graczyk2014}) to derive $M_{\rm bol}$  ($M_{\rm bol}=$ $m_{K} - \mu - A_K + $BC$_K$). Using the magnitude-luminosity relation, we then derived L$_*$ ($L = L_0 \times  10^{(M_{\rm bol})/2.5}$, with $L_0$ being the zero point luminosity of the sun $ \sim 3\times 10^{28}\,\ \textrm W$). The radius (R$_*$) was derived from the Stefan-Boltzmann equation ($L=4\pi \sigma R^2 T^4_{\rm eff}$). The parameters corresponding to the best fit model of each star (i.e. $\chi^2_{\rm red,min}$, $T_{\rm eff,TiO}$, $A_V$, $M_{\rm bol}$, log(L$_*$/L$_{\odot}$) and R$_*$) are presented in Table \ref{ResultsTable}. \\
\indent
To propagate the uncertainties, similar to \cite{Davies2010}, we constructed a 2D $\chi^2$ map for all fitted stars (see Fig. \ref{Chisq16}). The $\chi^2$ map shows the derived $\chi^2$ for every point on the 2D grid. Using a $\chi^2_\textrm {min} + \Delta\chi^2$ (where $\Delta\chi^2 = 2.3$, 6.2 and 11.8 for 1, 2 and 3$\sigma$), we were able to derive uncertainties from the contours around the best fit solution and take the extreme points as the inferred uncertainties. Following standard error propagation rules, we then estimated uncertainties for $M_\textrm {bol}$, log(L$_*$/L$_{\odot}$) and R$_*$ and listed the uncertainties in Table \ref{ResultsTable}.

\begin{table*}
\centering     
  \begin{threeparttable}
    \caption{Parameters derived for stars of spectral class M.}
    \tiny
    \label{ResultsTable}
    \renewcommand{\arraystretch}{1.3}
        \begin{tabular}{l r l l l l l c}        
        \hline\hline                 
        ID & $\chi^2_{\rm red}$ & $T_{\rm eff,TiO}$ & $A_V$ & $\rm log\,\it g$ & $M_\textrm{bol}$ & log(L$_*$/L$_{\odot}$)& R$_*$\\    
         &  & (K) & (mag) & (dex) & (mag) & (dex) & (R$_{\odot}$)\\         
        \hline\hline                        
            SMC3 & 1.0 & 3570$^{+21}_{-32}$   & 0.15$^{+0.04}_{-0.08}$ & +0.3 $\pm$ 0.1 & -7.50$^{+0.20}_{-0.21}$ & 4.90$^{+0.08}_{-0.08}$ & 730$^{+75}_{-65}$ \\ 
            LMC1 & 1.0 & 3600$^{+68}_{-37}$   & 0.40$^{+0.13}_{-0.12}$ & +0.4 $\pm$ 0.1 & -7.92$^{+0.15}_{-0.15}$ & 5.06$^{+0.06}_{-0.06}$ & 875$^{+70}_{-60}$ \\ 
            LMC2 & 5.2 & 3410$^{+32}_{-52}$   & 0.30$^{+0.20}_{-0.19}$ & > +1.0 & -6.90$^{+0.23}_{-0.22}$ & 4.65$^{+0.09}_{-0.09}$ & 610$^{+75}_{-60}$ \\ 
            LMC3 & 2.9 & 3570$^{+59}_{-46}$   & 1.00$^{+0.14}_{-0.14}$ & +0.5 $\pm$ 0.1 & -8.89$^{+0.18}_{-0.18}$ & 5.45$^{+0.07}_{-0.07}$ & 1390$^{+130}_{-110}$ \\ 
            LMC4 & 1.1 & 3710$^{+216}_{-87}$  & 0.90$^{+0.18}_{-0.18}$ & +0.1 $\pm$ 0.1 & -8.32$^{+0.21}_{-0.21}$ & 5.22$^{+0.08}_{-0.08}$ & 990$^{+115}_{-100}$  \\ 
            LMC5 & 1.6 & 3620$^{+53}_{-55}$   & 0.35$^{+0.10}_{-0.13}$ & +0.3 $\pm$ 0.1 & -7.82$^{+0.13}_{-0.15}$ & 5.03$^{+0.05}_{-0.06}$ & 825 $^{+60}_{-60}$  \\ 
            LMC6 & 1.9 & 3640$^{+59}_{-51}$   & 0.35$^{+0.08}_{-0.14}$ & > +0.5 & -8.01$^{+0.11}_{-0.16}$ & 5.10$^{+0.05}_{-0.06}$ & 890 $^{+55}_{-65}$  \\ 
            LMC7 & 3.5 & 3600$^{+25}_{-49}$   & 0.50$^{+0.05}_{-0.14}$ & > +0.5 & -7.94$^{+0.10}_{-0.16}$ & 5.07$^{+0.04}_{-0.06}$ & 880 $^{+45}_{-65}$   \\ 
            LMC8 & 2.5 & 3540$^{+17}_{-48}$   & 0.30$^{+0.05}_{-0.18}$ & +0.5 $\pm$ 0.1 & -7.61$^{+0.12}_{-0.21}$ & 4.94$^{+0.05}_{-0.08}$ & 780 $^{+50}_{-70}$ \\ 
        \hline
        \end{tabular}
    \begin{tablenotes}
        \small
        \vspace{5pt}
        \item \textit{Notes}: The errors presented for $T_{\rm eff,TiO}$ and $A_V$ correspond to the edges of the 1-$\sigma$ contour. For LMC6 and LMC7, we give a lower limit on the surface gravity. By eye we noticed that +0.5 came close to fitting these stars, but the real surface gravity is likely in the range from [+0.5,+1.0], which could not be fitted with the models we have available. For LMC2, it is evident that the surface gravity is higher than +1.0, and therefore too compact to be a RSG.
    \end{tablenotes}
    
  \end{threeparttable}
\end{table*}

\subsection{Comparison to $T_{\rm eff}$($J-K$)}
The effective temperature one derives depends on the region in the RSG atmosphere probed, resulting in systematic offsets between different methods. Numerous approaches have been developed in the last two decades to derive $T_{\rm eff}$ for RSGs. Each of these approaches uses a different set of spectral or photometric features, so that depending on the data available, for both optical and near-IR wavelengths, methods have been established to derive $T_{\rm eff}$ for RSGs. One approach is to convert photometric data into an effective temperature through empirical relations. As near-IR wavelengths are not very susceptible to extinction and variability \citep{Loon1999,Whitelock2003,Mauron2011} we opted to use the $T_{\rm eff}$($J-K$) relations \citep{Tabernero2018,Britavskiy2019b,Britavskiy2019a}. These relations probe a temperature closer to the $\tau_\lambda = 2/3$ continuum region, allowing us to discuss potential weaknesses of the TiO method as the molecular absorption happens in layers where $\tau_\lambda \leq 2/3$. \cite{Britavskiy2019b,Britavskiy2019a} fitted a sample of several hundreds of RSGs from \cite{Tabernero2018}. In this study, \cite{Tabernero2018} derived effective temperatures from spectral lines in the $8400-8800\,\textrm{\AA}$ region by averaging the $T_{\rm eff}$ derived from plane-parallel Kurucz models \citep{Meszaros2012} and spherical \textsc{marcs} models. \cite{Britavskiy2019b,Britavskiy2019a} then obtained a linear relation between the $T_{\rm eff}$ and the ($J-K$) colour for both Magellanic Clouds. However, $T_{\rm eff}$($J-K$) does not serve as a direct comparison to the $T_{\rm eff,TiO}$ due to the nature of the derived spectroscopic $T_{\rm eff}$. A direct comparison could only be made if a relation between the $J-K$ colours and the $T_{\rm eff,TiO}$ were to be established in the future, or if one would construct a relation between the $J-K$ colours and the temperature of the \textsc{marcs} models, which will be attempted in a future work. \\
\indent
We used the relations presented in \cite{Britavskiy2019b,Britavskiy2019a}, which are applicable for the LMC and SMC, respectively;

\begin{ceqn}
\begin{align}
    \rm T_{eff} (\it J-K) \rm  = -791 \cdot (\it J-K\rm _S)_0 + 4741
    \label{Eq2}
\end{align}
\end{ceqn}

\begin{ceqn}
\begin{align}
    \rm T_{eff} (\it J-K) \rm = -1432 \cdot (\it J-K\rm _S)_0 + 5549
    \label{Eq3}
\end{align}
\end{ceqn}
\noindent
First we correct for the reddening by calculating $(J-K_S)_0$ from the relation $(J-K_S)_0=(J-K_S)-E(J-K)$, where $E(J-K)=0.535E(B-V)$ \citep{Schlegel1998} and $E(B-V)= \frac{A_V}{R_V}$, with $R_V=2.74$ and $R_V=3.41$ for the SMC and LMC, respectively, so that ultimately the reddening correction depends on the derived $A_V$ from Section \ref{Sec3.1}. Although this affects the integrity of the comparison, as we use our derived $A_V$ to deredden the independent $J-K$ colour, using an independent $A_V$ (i.e. from an extinction map) would not affect the $T_{\rm eff}$($J-K$) much ($\pm50\,\textrm{K}$ per 0.5~mag, which is arguably small compared to the adopted error of $\pm140\,\textrm{K}$). The resulting $T_{\rm eff}$($J-K$) using Eq. \ref{Eq2} $\&$ \ref{Eq3} are presented in the last column of Table \ref{TeffTable}. In most cases, $T_{\rm eff}$($J-K$) is higher than $T_{\rm eff,TiO}$, as expected. It is noted that Eq. \ref{Eq2} $\&$ \ref{Eq3} are valid between $0.8\leq J-K_S \leq1.4 \,\ \textrm {mag}$, yet two of our objects had colours slightly beyond the upper limit, so we extrapolated the relations accordingly. \cite{Britavskiy2019b,Britavskiy2019a} indicated a typical error of $\pm140\,\textrm{K}$ as the statistical uncertainty on their linear fit, which we adopted.
\begin{table}
\centering     
  \begin{threeparttable}
    \caption{Spectroscopic and photometric $T_{\rm eff}$ for the M supergiants.}
    \tiny
    \label{TeffTable}
    \renewcommand{\arraystretch}{1.3}
        \begin{tabular}{l l l}        
        \hline\hline                 
        ID & $T_{\rm eff,TiO}$ & $T_{\rm eff}$($J-K$) \\ 
         & (K) & (K) \\    
        \hline\hline                        
            SMC3 &  3570$^{+21}_{-32}$  & 4020 $\pm$ 140 \\
            LMC1 &  3600$^{+68}_{-37}$  & 3685 $\pm$ 140 \\
            LMC3 &  3570$^{+59}_{-46}$  & 3630 $\pm$ 140 \\
            LMC4 &  3710$^{+216}_{-87}$  & 3685 $\pm$ 140 \\
            LMC5 &  3620$^{+53}_{-55}$  & 3830 $\pm$ 140 \\
            LMC6 &  3640$^{+59}_{-51}$  & 3880 $\pm$ 140 \\
            LMC7 &  3600$^{+25}_{-49}$  & 3795 $\pm$ 140 \\
            LMC8 &  3540$^{+17}_{-48}$  & 3805 $\pm$ 140 \\

        \hline
        \end{tabular}
    \end{threeparttable}
\end{table}

\section{Discussion} \label{Sec4}
\subsection{Hertzsprung-Russell Diagram analysis} \label{Sec4.1}

We have used evolutionary models to verify the evolutionary status through the derived $T_{\rm eff}$ and $L$ of our stars. We use MIST models \citep{Dotter2016mist,Choi2016mist} for rotating stars ($v/v_{crit}=0.4$) in the range of 8 to $25\,\textrm{M}_{\odot}$. We proceeded with the MIST models over the Geneva models, as they better reproduced the RSG population from \cite{Yang2021} at higher masses. The goal of this section is to provide a crude comparison of our sample to the tracks, not to derive properties. The tracks terminate at carbon core depletion, thus approaching the imminent collapse as a supernova, effectively mapping the entire evolution of a massive star from the main-sequence to the final evolutionary stages.

\begin{figure*}
    \centering
    \includegraphics[width=1.3\columnwidth]{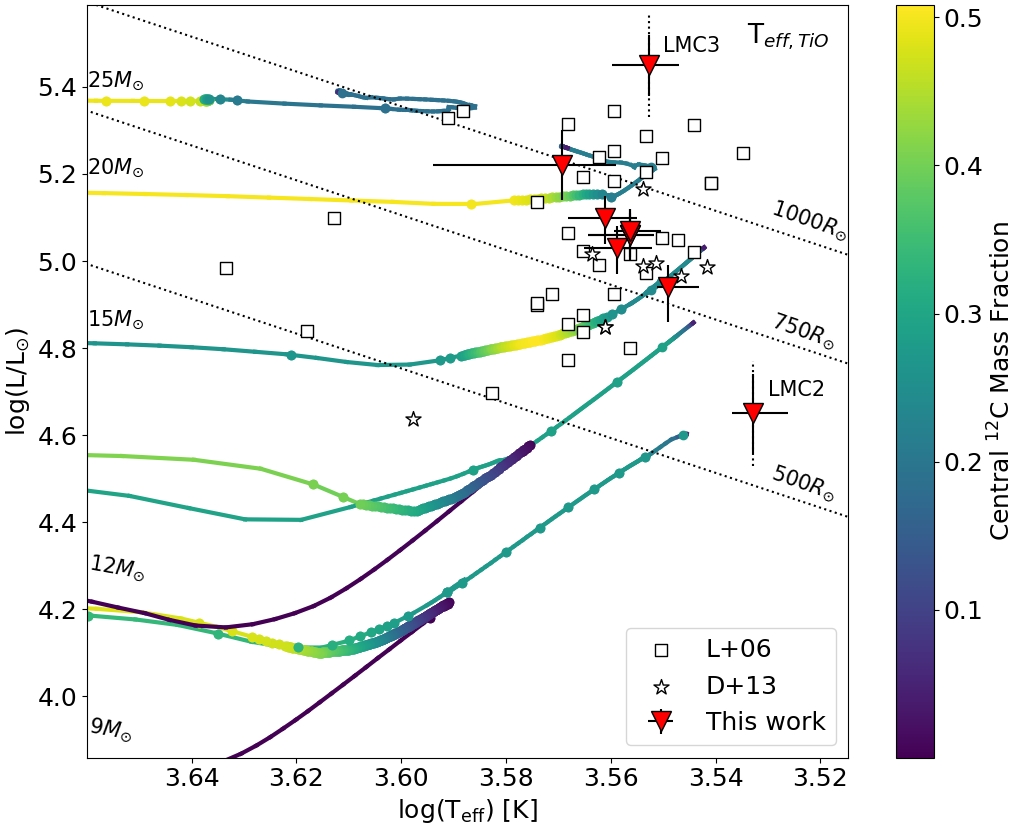}
    \\
    \includegraphics[width=1.3\columnwidth]{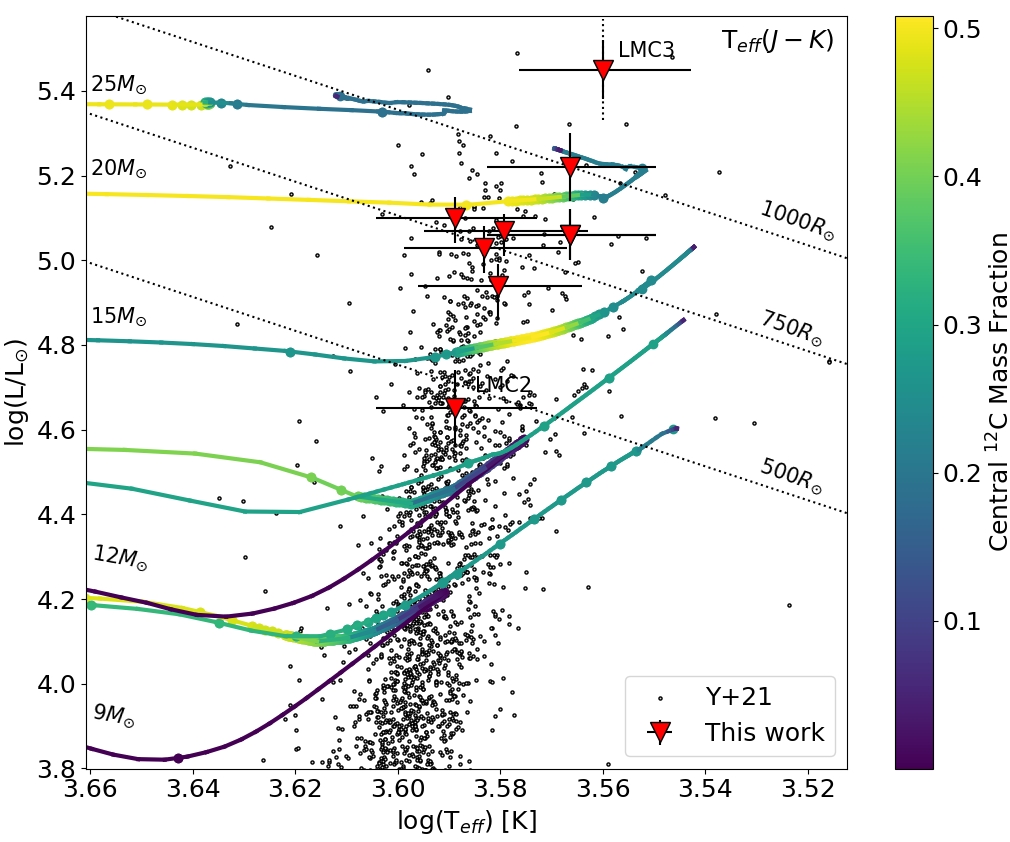}
    \caption{Top: HRD indicating the locations of our LMC targets with inverted red triangles. The $T_{\rm eff}$ for all data points was derived through the TiO method. Smaller light grey squares and stars are objects from \cite{Levesque2006} and \cite{Davies2013}, respectively. For the two outliers we have extended the uncertainty assuming a shift of 0.3~mag in the $K-$band (dotted vertical error bar) instead of only the propagated uncertainty, to visualize the effect of intrinsic variability. The colour map represents the central ${^{12}{\rm C}}$ mass fraction, while the nodes on the track again indicate a step of $10^4$ years. Bottom: Same as top, but for $T_{\rm eff}$ derived using the $T_{\rm eff}$($J-K$) method. We show the general RSG population in the background \cite[][black points]{Yang2021}.}
    \label{HRDs}
\end{figure*}

\indent
We overlay evolutionary tracks on our data in Fig. \ref{HRDs}. We also added sources from \cite{Levesque2006} and \cite{Davies2013}, who presented similarly derived $T_{\rm eff,TiO}$ results, allowing us to make a direct comparison between the two samples. We note that a considerable fraction of the RSGs lie in the forbidden zone of the evolutionary tracks right of the Hayashi limit. Recall that the $T_{\rm eff}$ of these points is determined from the TiO bands, while the $T_{\rm eff}$ of the models probe the (hotter) continuum temperature. Indeed, when we construct a HRD using the $T_{\rm eff}$($J-K$) approach (see Fig. \ref{HRDs}, right panel), we see that the discrepancy becomes minimal and only a few outliers remain. Binary interactions cannot be used to explain the position in the HRD for these outliers, as the location of the RSGs in the HRD is largely unaffected by the amount of envelope mass of the RSGs, except for very low or very high envelope masses \citep{Farrell2020,Beasor2020}, it is difficult to infer these scenarios. For moderate envelope masses, an increase in envelope mass would make an object appear slightly hotter.\\
\indent
Most of our objects are in agreement with predictions from single star evolutionary tracks. We identified three potential outliers: LMC2, LMC3 and SMC3. \\
\indent
\textit{i}) LMC2 (M4--5~II--III) is located at the lower luminosity end of the population compared to the RSGs studied in \cite{Levesque2006,Davies2013}. This is in agreement with the luminosity classification we have assigned. Using the derived $T_{\rm eff,TiO}$, it appears that LMC2 is too cool to be explained by any of the evolutionary tracks (see Fig. \ref{HRDs}, left panel). This source is well within the range of tracks if we consider $T_{\rm eff}$($J-K$) however (see right panel of Fig. \ref{HRDs}). Despite the spectral classification, and considering that the $T_{\rm eff}$($J-K$) is closer to the real temperature of the photosphere, we find that the properties of this star are consistent with RSGs, which is in conflict with the spectral classification. It is possible however that there is a circumstellar contribution to the $K-$band magnitude, which results in a higher luminosity. \\
\indent
\textit{ii}) LMC3 (M3~I) is brighter than other M-type supergiants. Since the star appears to be saturated in archival optical HST images, we could not find direct evidence of a contaminated $K-$band magnitude by an unresolved binary, which could potentially affect the derived luminosity. If this star is truly single, it may be grouped amongst the largest supergiants ($\rm R=1390\rm \,\ R_{\odot}$) known in the LMC, albeit still trailing WOH G64 in size ($\rm R=1540\rm \,\ R_{\odot}$, log(L$_*$/L$_{\odot})\sim$5.45; \citealt{Levesque2009}). Even though LMC3 is slightly brighter than WOH G64, we note that the luminosity of WOH G64 has been carefully corrected for contaminating contributions in the $K-$band magnitude, while LMC3 is not. Further investigation is needed to carefully constrain the stellar radius and luminosity of LMC3 with more certainty. Considering the RSG sample of \citet{Davies2018}, this object is at the high luminosity end of what has been observed and what is predicted by the Geneva models at Z = 0.006 from \citet{Eggenberger2021}, indicating that this star is a high mass RSG at the latest stages of evolution. \\
\indent
\textit{iii}) SMC3 (M2~I) is one of the latest type RSGs in the SMC and lies in the forbidden zone as can be seen in the left panel of Fig. \ref{HRDs2}. SMC3 does not cluster with the sources from similar studies and appears to be cooler and less luminous. However, if we employ a different methodology to derive the effective temperature (i.e. $T_{\rm eff}$($J-K$), see right panel of Fig. \ref{HRDs2}), this source is not an outlier. We also note that if one considers the variability of the stars, LMC2 and SMC3 may not strictly be outliers.

\subsection{Metallicity dependence}
As stated in Section \ref{Sec3.1}, we are guided by empirically derived metallicities from \citet{Davies2015} (log$(Z/Z_{\odot})=-0.35\,\textrm{dex}$ for the LMC and log$(Z/Z_{\odot})=-0.55\,\textrm{dex}$ for the SMC RSG populations, respectively) and we proceeded with the spectral modelling using the most appropriate models available for these values. Given that \textit{i}) the models and empirical values are slightly different, \textit{ii}) we have fitted metal-dependent molecular bands and \textit{iii}) the metallicity may not strictly be uniform within the Magellanic Clouds, we tested if the metallicity could possibly affect the results. We argue that the use of one single metallicity per galaxy is justified: \textit{i}) The general metallicity gradient ([Fe/H] abundance) in the SMC is flat and shows a small linear decrease outwards for the LMC \citep{Cioni2009}. \textit{ii}) \cite{Davies2013} discussed that small variations in metallicity do not affect the results significantly. Indeed, changing the value of log$(Z/Z_{\odot})$ significantly (by $\pm$ $0.25\,\textrm{dex}$), did not affect the effective temperature on a $50\,\textrm{K}$ scale in their study, and is therefore not expected to affect our results much.\\

\subsection{$T_{\rm eff,TiO}$ - $T_{\rm eff}$($J-K$) discrepancy}
The $T_{\rm eff,TiO}$ is generally systematically lower in comparison to continuum temperature methods (i.e. $T_{\rm eff}$($J-K$) and the $T_{\rm eff}$ derived through SED fitting), due to the region at which these TiO molecules are formed. Similar to $T_{\rm eff}$, the extinction $A_V$ is remarkably low. By selecting dusty, red sources in the CMDs presented in Section \ref{Sec2}, we expected to find high extinction objects. Using this fitting method however, we find extinction factors $A_V < 1$, despite the fact that the extinction should be the sum of the contributions from foreground extinction, extinction from the star forming region and circumstellar extinction. \\
\indent
The SED method has been indicated by many authors to be a more reliable method to extract $T_{\rm eff}$ compared to the TiO bands \citep{Davies2013,Gonzalez2021,DaviesPlez2021}. This is due to the fact that the TiO molecular absorption happens at a region in which $\tau_\lambda \neq 2/3$, not near the photosphere. Here, we stress that we have derived the $T_{\rm eff,TiO}$, not the true $T_{\rm eff}$ of the RSG, which is technically defined to be at $\tau_\lambda = 2/3$. In \cite{Davies2013} it has been discussed extensively that the \textsc{marcs} models do not correctly represent the radial temperature structure of a real RSG, as the line formation zone of the TiO bands is pushed to higher altitudes and lower temperatures. To fit the optical molecular bands properly, detailed 3D models that correctly take into account convection are needed. \cite{Davies2013} conclude that the temperature of the layer at which the continuum forms is consistent between 1 and 3D models, hence they argue that, for 1D models, using the continuum is more reliable. Finally, they suggest that the TiO method may be further hampered by the fact that TiO molecules may form in the wind region and the TiO absorption strengths therefore also depend on the wind properties. \\
\indent
Some studies however, provide opposing results. The recent study by \cite{Massey2021} suggested that Geneva single star \citep{Eggenberger2008,Ekstrom2012} and BPASS binary \citep{BPASS1} models fit observed RSG/WR ratios the best if the adopted effective temperature is the one derived through spectroscopy, not photometry (i.e. $T_{\rm eff}$($J-K$)). They argue that using the spectroscopic derived $T_{\rm eff}$ of stars yields a better match of their position in the HRD with respect to evolutionary tracks. However, \cite{Massey2021} also mention that it is uncertain whether the $T_{\rm eff}$ adopted in the models can be interpreted as equal to what is derived from a real RSG spectrum. 
%

\section{Summary and conclusions} \label{Sec5}
We set out to find dusty, luminous evolved massive stars using near and mid-IR selection criteria. We have spectroscopically confirmed eight new dusty, luminous RSGs, a new bright giant (LMC2) and a new YSG (SMC2, with strong H$\alpha$ emission) and three contaminants (i.e. two giants and a LBVc + H\textsc{ii} region) through the analysis of newly obtained MagE spectra. In addition, we found a different spectral type for the sgB[e] (SMC1, A0 I[e]), compared to the previous classifications reported in the literature (B8~I[e] and A1~I[e]), suggesting this star has changed spectral type more than once over a thirty year period, although we remain speculative over the physical mechanism modifying the photosphere of the central star.
The RSGs were analysed and modelled using the \textsc{marcs} models to derive key properties. We approached the modelling by fitting the optical TiO bands to a precision of $50\,\textrm{K}$, and have used evolutionary tracks to determine the current evolutionary stage of these objects. We have derived the physical properties of these red supergiants. Most sources are very luminous (log(L$_*$/L$_{\odot}) \geq 5 \textrm {\,\ dex}$) and have large radii (R$_* \sim 700-1400$ R$_{\odot}$). The brightest RSG (LMC3) appears disconnected from the bulk of the RSG population in the HRD. This object is located at the tail of the empirical and theoretical RSG luminosity distribution and is arguably one of the brightest and largest stars known in the LMC, with log(L$_*$/L$_{\odot}) \sim 5.45 \textrm {\,\ dex}$ and R$_* \sim 1400$ R$_{\odot}$, resembling WOH G64 in size and luminosity \citep{Levesque2009}.

We have compared our results and methodology with respect to other studies, computed the $T_{\rm eff}$($J-K$) of our RSGs and discussed the implications. The majority of our RSGs have a $T_{\rm eff,TiO}$ that is arguably different from what is expected through $T_{\rm eff}$($J-K$), although hotter temperatures are expected for $T_{\rm eff}$($J-K$) as this temperature probes a deeper layer, closer to $\tau_\lambda \sim 2/3$.

Despite the aim to select dusty supergiants, using the methodology as presented in this paper has resulted in extinction coefficients ($A_V < 1$~mag) lower than expected. \cite{Davies2013} demonstrated that the derived extinction from the optical was in almost all cases much lower than that derived using the near-IR. We share their conclusion that using the optical TiO bands to derive the extinction is not ideal. We suggest the use of near-IR spectra or SED fitting for a more reliable result. 

We conclude that the selection criteria were successful in selecting cool, luminous RSGs. However, in the near future we aim to apply the machine learning algorithm by \cite{Maravelias2022} to improve the selection of candidate evolved massive stars. The next step is to employ DUSTY models \citep{Dusty1} to derive mass loss rates from the SED (e.g.\ Yang et al. 2022, in prep.). The derivation of the central star properties constrains the parameter space used with DUSTY. This significantly decreases the amount of viable models, which reduces degeneracies and computation time. Finally, we emphasize that LMC3 is a remarkable source and we aim to present a more detailed analysis of this star in a future study.

\section*{Acknowledgements}
We thank the referee, Ben Davies, for the careful reading of the manuscript and the insightful comments and suggestions that significantly improved the manuscript. SdW, AZB, FT, GM, MY, EZ acknowledge funding support from the European Research Council (ERC) under the European Union’s Horizon 2020 research and innovation program (Grant agreement No. 772086). NB acknowledges support from the postdoctoral program (IPD-STEMA) of Liege University. EZ also acknowledges support by the Swiss National Science Foundation Professorship grant (project number PP00P2 176868; PI Tassos Fragos). This paper includes data gathered with the 6.5 meter Magellan Telescopes located at Las Campanas Observatory, Chile. This research has made use of NASA's Astrophysics Data System. This research has made use of the SIMBAD database, operated at CDS, Strasbourg, France. This research has made use of the VizieR catalogue access tool, CDS, Strasbourg, France. 

\bibliographystyle{aa} 
\bibliography{bib.bib} 

\appendix
\section{Supplementary Figures} \label{App1}

\begin{figure*}
\begin{center}
    \centerline{\includegraphics[width=2.1\columnwidth]{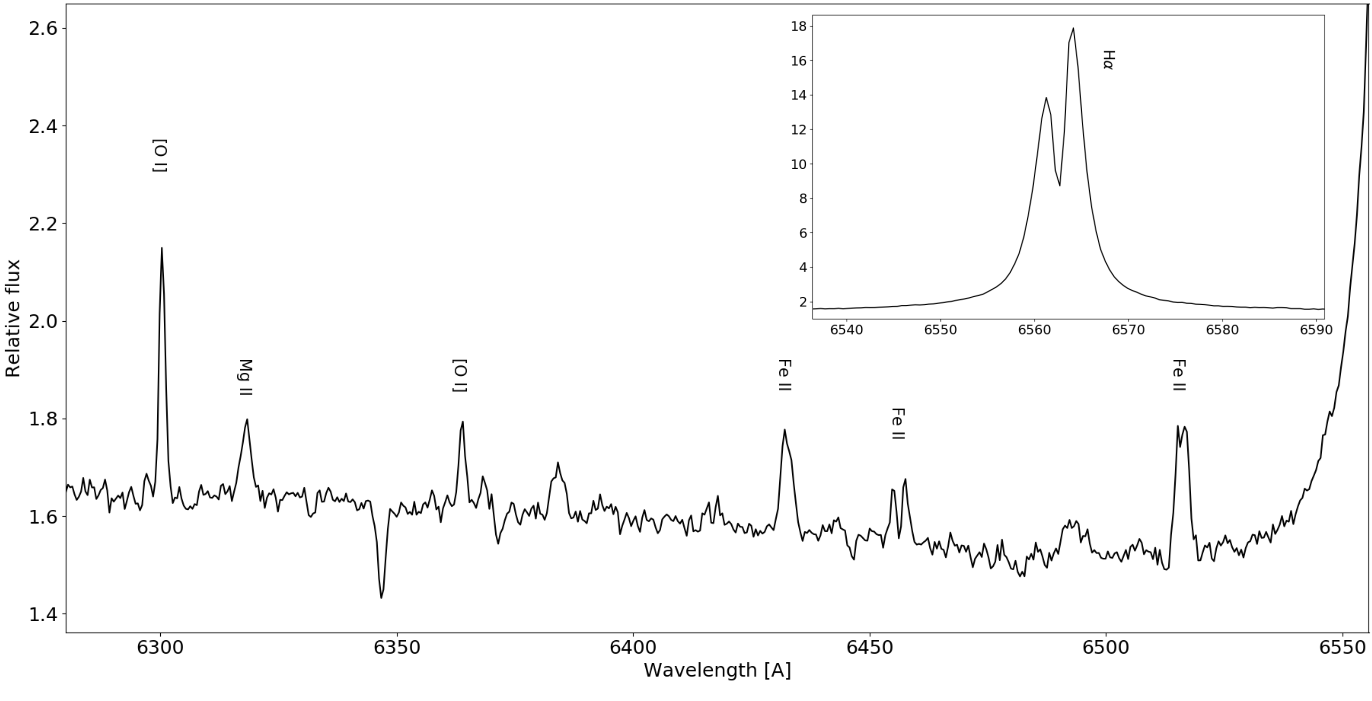}} 
    \caption{Prominent disk emission lines of SMC1 near the double peaked H$\alpha$ emission, which is shown in the inset at the top right.}
    \label{LHA115S23Z1}
\end{center}
\end{figure*}

\begin{figure*}
\begin{center}
    \centerline{\includegraphics[width=2.1\columnwidth]{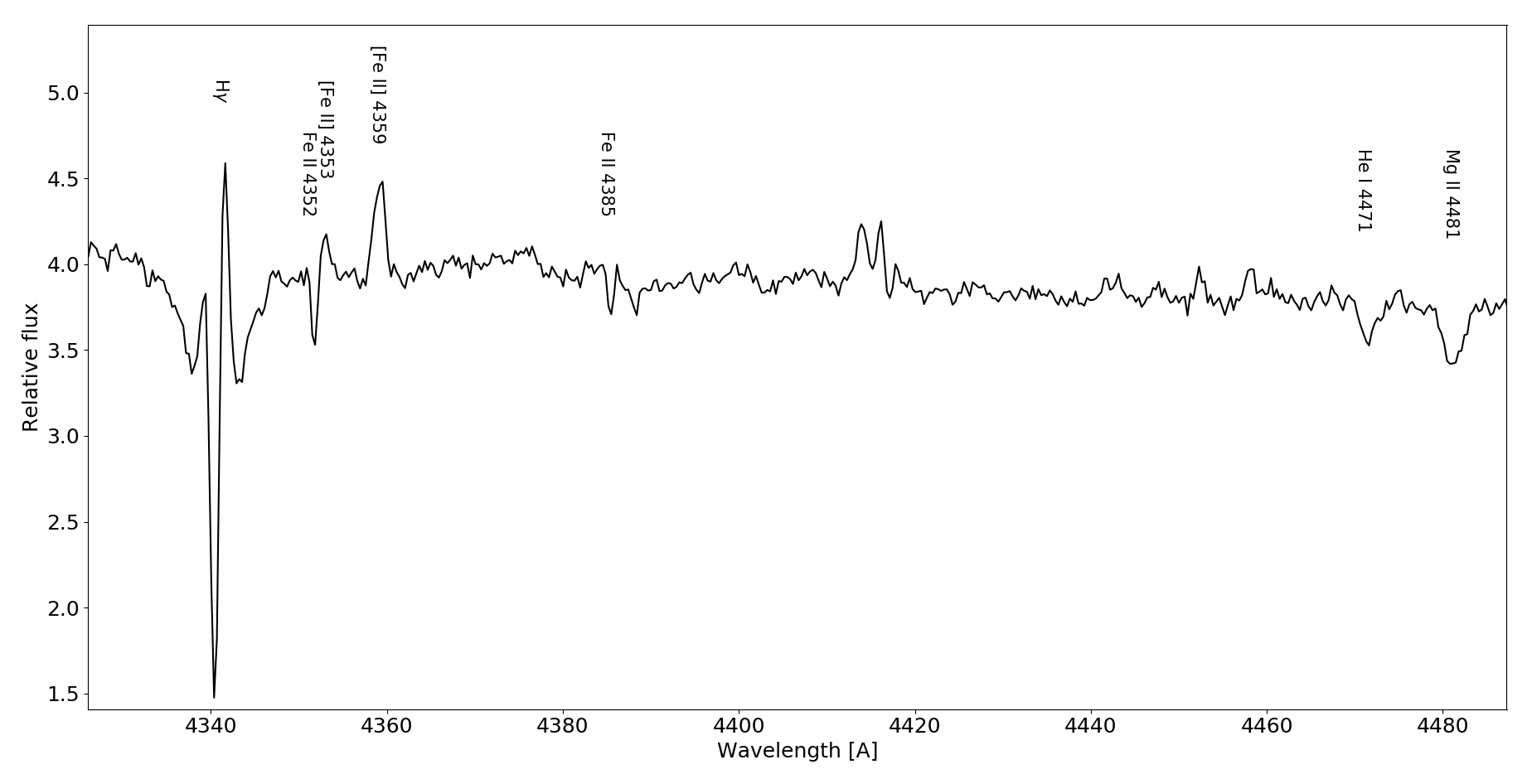}} 
    \caption{Zoom in on the metal lines used for the A0~I[e] spectral classification of SMC1.}
    \label{LHA115S23Z2}
\end{center}
\end{figure*}

\begin{figure*}
\begin{center}
    \centerline{\includegraphics[width=2.1\columnwidth]{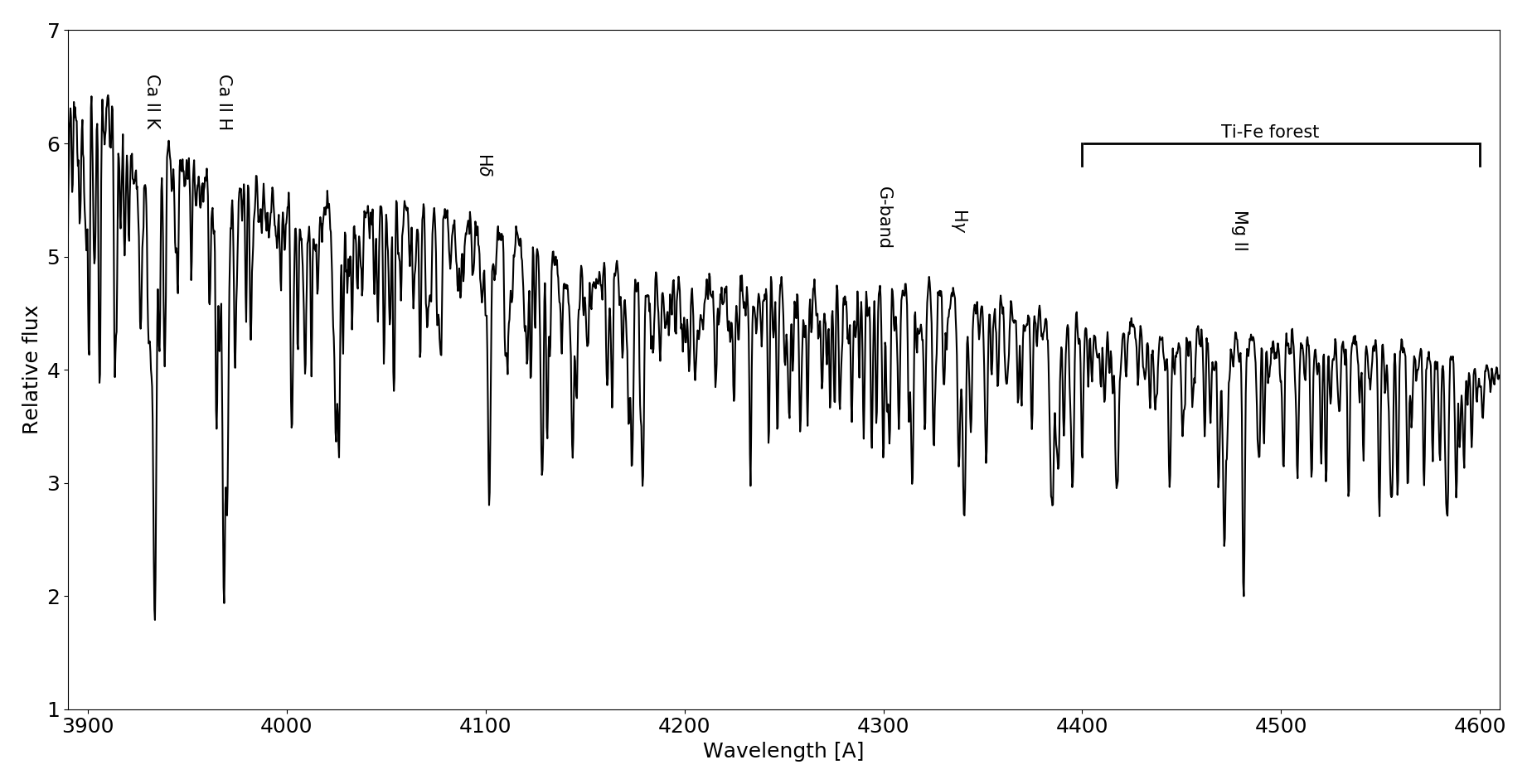}} 
    \caption{Zoom in on specific lines selected for the classification for SMC2 (spectral type F8~I). Balmer lines are moderately strong, metal lines are abundant, and the G-band is weak.}
    \label{MA93Z1}
\end{center}
\end{figure*}

\begin{figure*}

    \begin{subfigure}{\textwidth}
        \includegraphics[width=1\columnwidth]{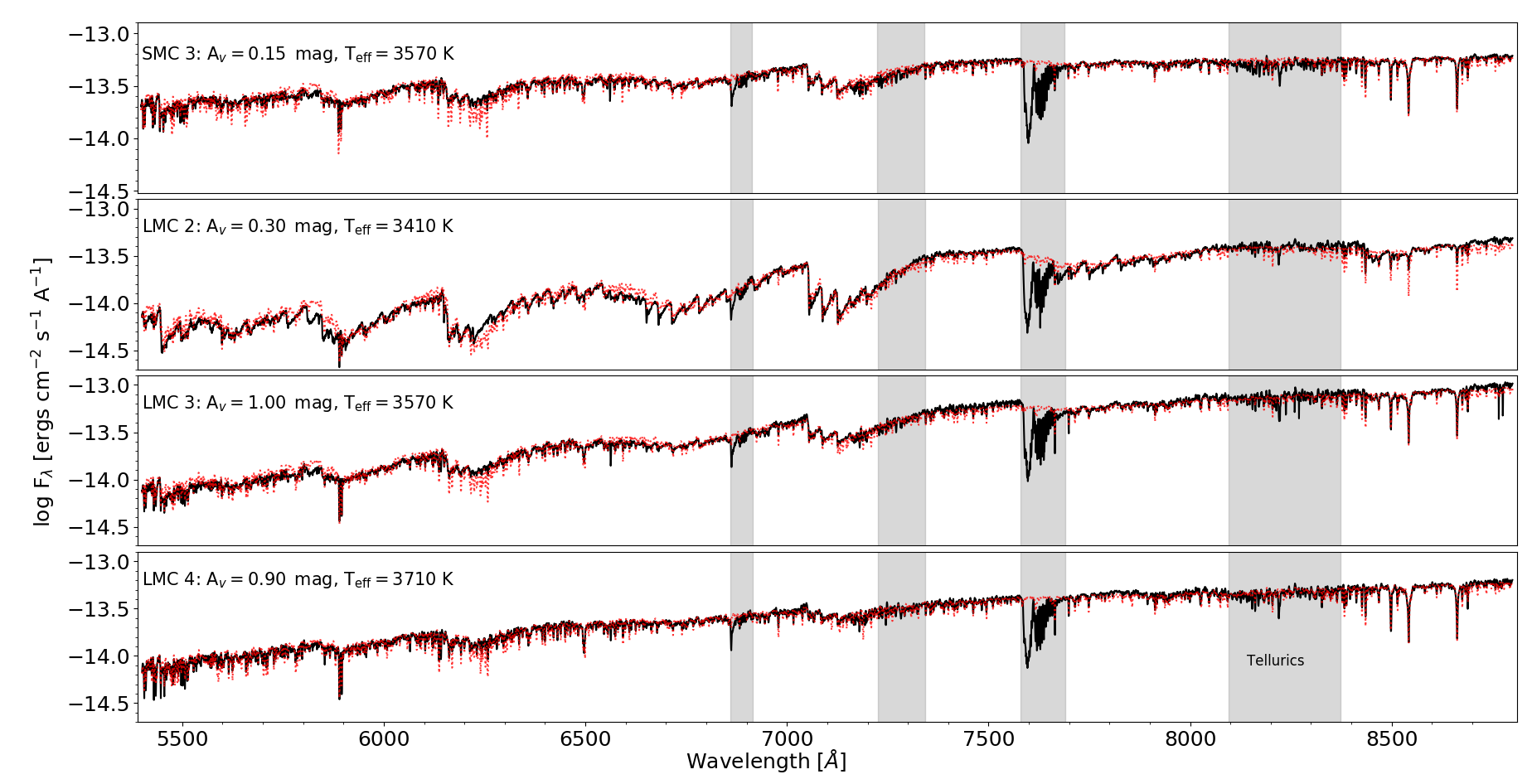}
    \end{subfigure}
\end{figure*}
\clearpage
\begin{figure*}
    \ContinuedFloat 
    \begin{subfigure}{\textwidth}
        \includegraphics[width=1\columnwidth]{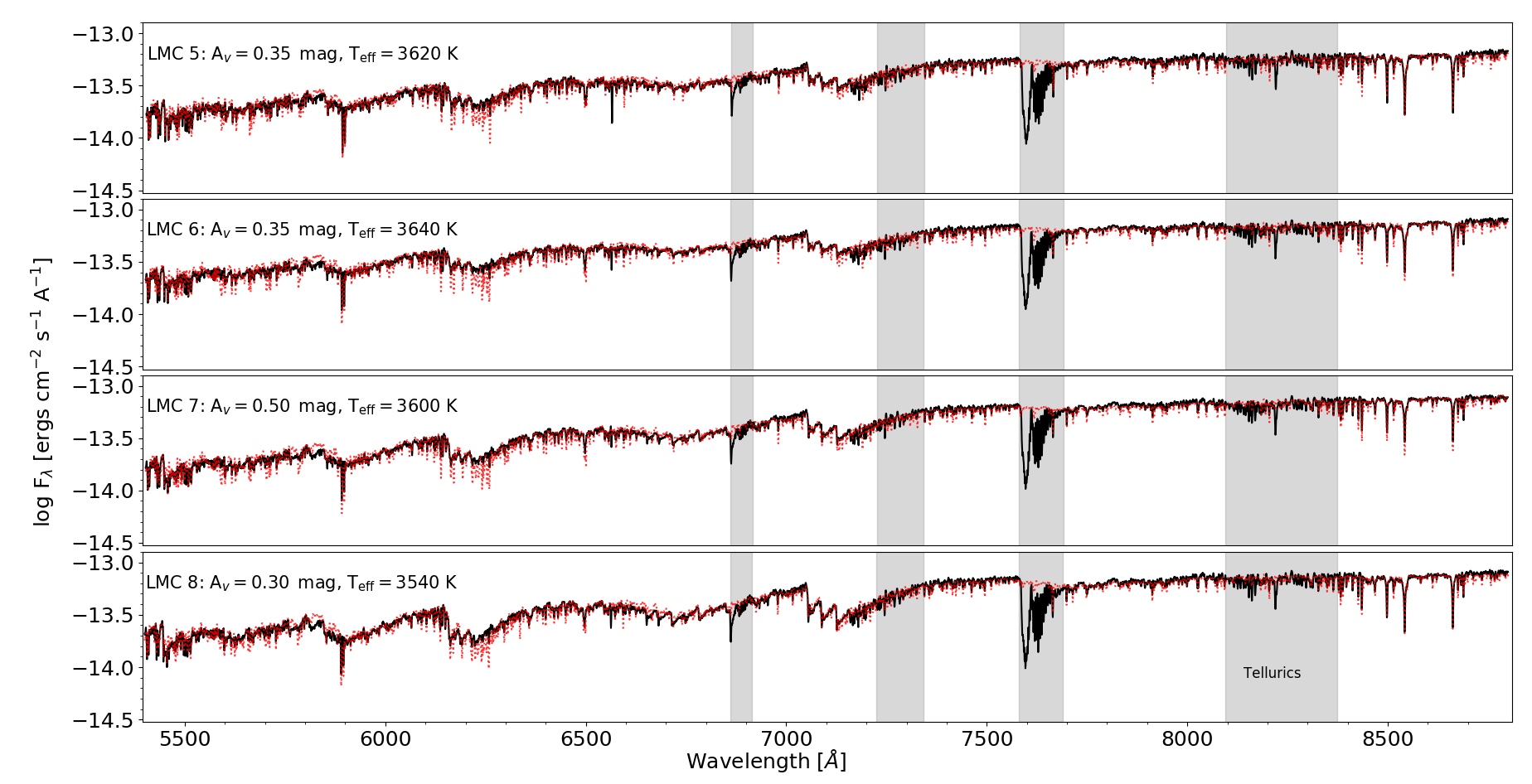}
    \end{subfigure}
\caption{Same as Fig.~\ref{BestFit16}, but for SMC3 and LMC2-8.}
\label{BestFitAll}  
\end{figure*}

\begin{figure*}
    \includegraphics[width=1\columnwidth]{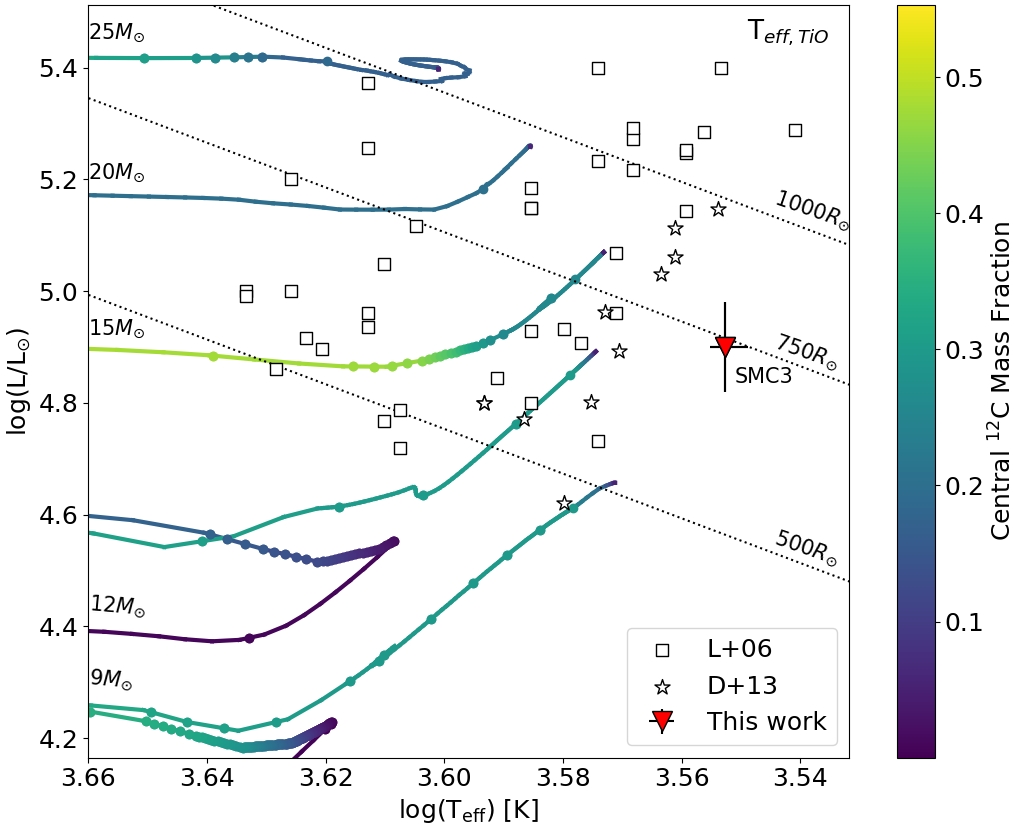}
    \includegraphics[width=1\columnwidth]{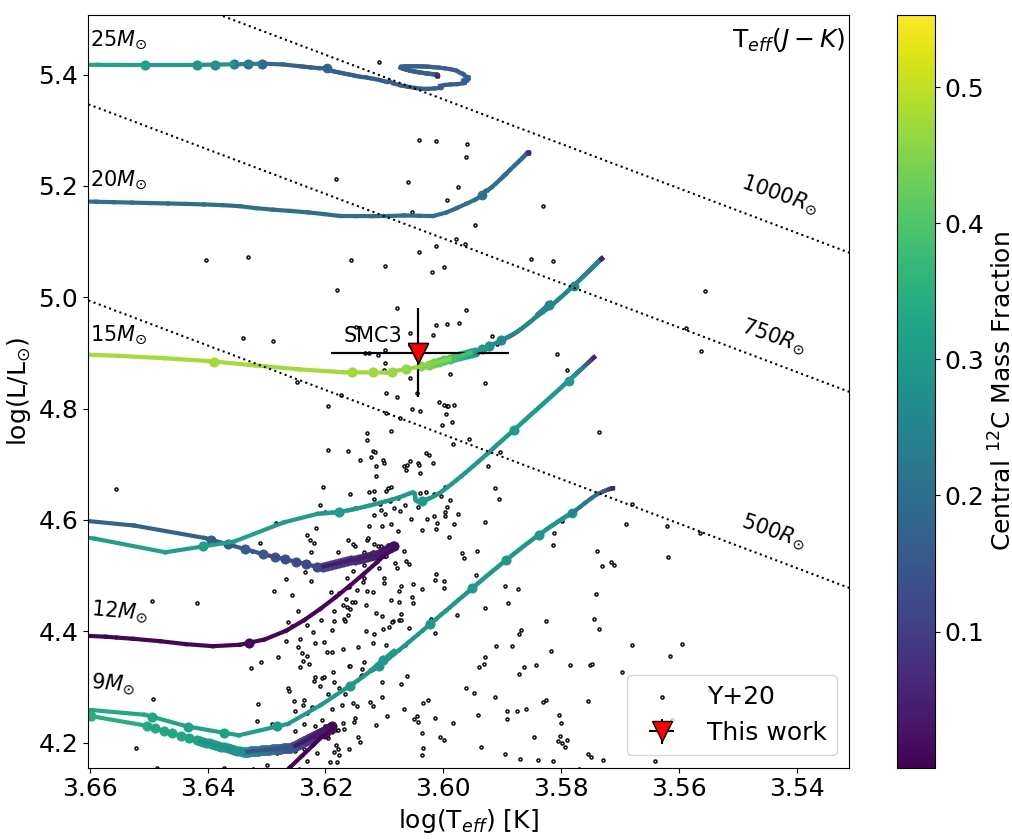}
    \caption{Left: same as top panel in Fig. \ref{HRDs}, but for the SMC. Right: same as bottom panel in Fig. \ref{HRDs}, but we show the general SMC RSG population in the background from \cite{Yang2020}.}
    \label{HRDs2}
\end{figure*}

\end{document}